\documentclass[twocolumn,superscriptaddress]{revtex4-2}
\usepackage{physics}
\usepackage{graphicx}
\usepackage{here}
\usepackage{amsmath}
\usepackage{amsthm}
\usepackage{amssymb}
\usepackage{bm}
\usepackage{wrapfig}
\usepackage{ascmac}
\usepackage{mathrsfs}
\usepackage{color}
\usepackage{comment}
\usepackage[normalem]{ulem}

\newcommand{\im}{{\rm i}}

\newtheorem{prop}{Proposition}
\newtheorem{asm}{Assumption}
\usepackage{comment}
\usepackage{hyperref}
\usepackage[normalem]{ulem}
\hypersetup{colorlinks=true,citecolor=blue,linkcolor=blue,urlcolor=blue}

\begin{document} 
\title{Time evolution of entanglement entropy after quenches in two-dimensional free fermion systems: a dimensional reduction treatment} 
\author{Shion Yamashika}
\affiliation{Department of Physics, Chuo University, Bunkyo, Tokyo 112-8551, Japan} 
\affiliation{SISSA and INFN, via Bonomea 265, 34136 Trieste, Italy}
\author{Filiberto Ares}
\affiliation{SISSA and INFN, via Bonomea 265, 34136 Trieste, Italy}
\author{Pasquale Calabrese}
\affiliation{SISSA and INFN, via Bonomea 265, 34136 Trieste, Italy}
\affiliation{International Centre for Theoretical Physics (ICTP), Strada Costiera 11,
34151 Trieste, Italy}
\date{\today}

\begin{abstract} 
We study the time evolution of the R\'enyi entanglement entropies following 
a quantum quench in a two-dimensional (2D) free-fermion system. 
By employing dimensional reduction, we effectively transform the 2D problem into decoupled chains, a technique applicable when the system exhibits translational invariance in one direction.
Various initial configurations are examined, revealing that the behavior of entanglement entropies can often be explained by adapting the one-dimensional quasiparticle picture.
However, intriguingly, for specific initial states the entanglement entropy saturates to a finite value without the reduced density matrix converging to a stationary state.
We discuss the conditions necessary for a stationary state to exist and delve into the necessary modifications to the quasiparticle picture when such a state is absent.
\end{abstract}

\maketitle

\section{Introduction} 

One of the most fundamental problems connecting quantum and statistical physics is  how a statistical ensemble emerges in a closed many-body quantum system that evolves unitarily~\cite{pssv-11, ge-16, cem-16}. 
The common wisdom is that entanglement generates non-local correlations unique to quantum systems that spread throughout the system by  unitary evolution: the resulting reduced density matrix for a local subsystem relaxes to a statistical ensemble such as the Gibbs ensemble~\cite{Deutsch-1991,Deutsch-1991,Srednicki-1994,Tasaki-1998,Rigol-2008, akpr-16}, or, in the case of integrable systems, a generalized Gibbs ensemble (GGE)~\cite{Rigol-2007,bs-08, cdeo-08,cef-12b, ilievski-15, vr-16, essler-2016}.  
Given that the (von Neumann) entanglement entropy quantifies the amount of entanglement between the subsystem and the rest of the system~\cite{amico-08, ccd-09, laflorencie-14}, understanding its behavior in a unitary evolution is important to clarify how entanglement spreads and how thermodynamics arises in isolated quantum systems. 
\par  
A popular and tractable setup to investigate the unitary evolution of a quantum many-body system is the quantum quench: the system is initially prepared in a non-equilibrium pure state $\ket{\psi_0}$ and then it is let evolve in time with a post-quench Hamiltonian $H$, $\ket{\psi(t)}=e^{-\im t H}\ket{\psi_0}$. 
In recent years, this protocol has been investigated not only theoretically but also experimentally thanks to remarkable developments in cold atom and ion trap systems \cite{Kinoshita-2006,hofferberth-2007,trotzky-2012,gring-2012,Cheneau-2012,langen-2013,mainert-2013, langen2015experimental,Islam-2015,Kaufman-2016}. 
\par 
The quench dynamics of the entanglement entropy in one-dimensional systems has been extensively studied in the literature. 
It has been found that it linearly increases in time and eventually saturates to a constant~\cite{Calabrese-2005}; the latter can be identified with the thermodynamic entropy of the (generalized) Gibbs ensemble  that describes locally the system at large times~\cite{dls-13,Alba-2017}. 
For integrable systems, this behavior is explained by the quasiparticle picture~\cite{Calabrese-2005} in which  the entanglement growth is due to the  propagation of pairs of entangled quasiparticles. 
This picture has been validated for one-dimensional free~\cite{Fagotti-2008, ep-08, eisler-2009, nr-14, bkc-14, bfsed-16, hbmr-18,ddr-23} and interacting integrable systems~\cite{Alba-2017, ac-18, mpc-19,ac-17a,mac-20,ckc-13} and generalized to different contexts~\cite{ac-20,ca-22,bfpc-18, abf-19, ma-20,kbp-21, bbtc-18, bc-18, bmc-20, bbc-20, kb-21, cara-22, acar-22, lcp-22,rc-23,ceh-19,ctd-19,ts-23,ge-19,fg-23,csrc-22,hcc-22, kkysytd-22, ykyt-22} and quantities~\cite{ac-23,cc-06, ac-19b, ac-19, ctc-14, mac-21, parez2021b, parez2021, mac-23, bcckr-22, bkccr-23, Ares-2023, amvc-23, ckacmb-23,makc-23,pbc-22,pvcc-22,bkalc-22,ghk-18,bkl-22,lsa-21,ggs-21,kkr-21,d-17}, see also the reviews \cite{c-18,c-20}. 
The same qualitative behavior for the time evolution of the entanglement entropy has been found in generic non-integrable/chaotic interacting models with no quasiparticles, see e.g.~\cite{lk-08, kh-13, pl-18, bkp-19, pbcp-20, nrvh-17, nz-20,mck-22}. 
For many years, there has been a prevailing belief that the microscopic mechanism for the entanglement growth is fundamentally different in integrable and chaotic systems; only recently a unifying picture emerged in the space-time duality approach~\cite{bkalc-22}.
The R\'enyi entanglement entropies are a natural and important generalization of von Neumann one, not only because they help to calculate the former via the replica trick~\cite{hlw-94,cc-04}, but also because they carry further relevant information about the system and are measurable in cold atom and ion trap experiments~\cite{Islam-2015, Kaufman-2016, lukin-2019, brydges-19, elben-2020, neven-21,vek-21,rvm-22}. 
While the quasiparticle picture captures the 
evolution of the R\'enyi entropies for free systems~\cite{Fagotti-2008,ac-17a}, it breaks down for interacting integrable models~\cite{bkalc-22}.

In higher dimensions $d\geq2$, the entanglement entropy in equilibrium has been largely investigated, see e.g. Refs.~\cite{ecp-10, pedc-05, gk-06, wolf-06, Li-06, bcs-06, cepd-06, ch-07, fz-07, Helling-11,jk-12, cmv-12,c-13, swingle, cmv-12b, aefsb-14, Frerot-2015, tr-19, mrc-20, fraenkel-20}; 
on the contrary its time evolution after a quantum quench has received little attention, mainly in field theory context~\cite{nnt-14, clm-16, cotler2016,lm-16,m-17,ms-20}. 
For this reason, here we study the quench dynamics of R\'enyi entanglement entropies in a 2D free-fermion system. In particular, we apply a dimensional reduction approach, which was introduced in Ref.~\cite{cp-00} and then has been applied to study, e.g., the (symmetry-resolved) entanglement entropy at equilibrium~\cite{aefsb-14, mrc-20}. 
For a finite 2D system, periodic in both directions, this approach works as follows. 
The initial configuration should be translationally invariant (non necessarily one-site shift invariant as we shall see) along one of the axes.
Next we should choose as subsystem a periodic strip in this direction, as shown in Fig.~\ref{fig:subsystem}. 
We can then decompose the R\'enyi entanglement entropies
into the sum of the single-interval entanglement entropies of decoupled one-dimensional (1D) systems, for which exact results are known.
We apply this strategy to calculate analytically the time evolution of the R\'enyi entanglement entropies for several particular initial configurations. 
We will see that our results can be explained in terms of a direct adaptation of the 1D quasiparticle picture, except for one particular initial configuration. 
The reason of such a mismatch is that the reduced density matrix does not attain a stationary value --- even if its entanglement entropy does tend to a constant value. 
We then discuss the general conditions under which there
is no stationary state in our 2D setting. 
From this observation, we deduce how the quasiparticle picture has to be modified to describe
the behavior of the entropy in the absence of a stationary state. 
\par 
The paper is organized as follows. In Sec.~\ref{sec:Setup}, we 
introduce the setup and some basic quantities, including the R\'enyi 
entanglement entropy. In Sec.~\ref{sec:dimension_reduction}, we 
describe the dimensional reduction approach. In Sec.~\ref{sec:examples}, we apply it to 
obtain analytically the behavior of the entropies in quenches from 
several initial configurations. In Sec.~\ref{sec:quasiparticle_picture}, we give a physical interpretation of the results obtained in the previous section in terms of the quasiparticle picture. 
In Sec.\,\ref{sec:reduced_density_matrix}, we analyze the conditions
for the existence of a stationary state and discuss how the quasiparticle picture modifies in that case. 
We finally draw our conclusions and present some outlooks in Sec.\,\ref{sec:summary}. 
We also include several appendices, where we derive  some of
the results presented in the main text.

\section{Setup and definitions}
\label{sec:Setup}

We consider free fermions on a 2D square lattice with isotropic hopping between nearest-neighbor sites. The system is described by the Hamiltonian 
\begin{align}
    H
    =
    -\frac{1}{2}
    \sum_{\bf \langle i,i'\rangle}
    a_{\bf i}^\dag a_{\bf i'}
    +\mathrm{H.c.},
    \label{eq:H}
\end{align}
where $\mathbf{i}=(i_x,i_y)$ is a vector identifying a site of the lattice, $\langle \mathbf{i,i'}\rangle$ stands for the nearest neighbors, and $a_\mathbf{i}=a_{i_x,i_y}(a_\mathbf{i}^\dag=a_{i_x,i_y}^\dag)$ is the annihilation (creation) operator of the fermion on the $\mathbf{i}$-th site. 
We assume that the system size $L_x(L_y)$ along the $x(y)$-axis is even and that periodic boundary conditions are imposed along both  directions. 
\par 
Moving to Fourier modes
\begin{align}\label{eq:b_modes}
    \tilde a_\mathbf{q}
    =
    \tilde a_{q_x,q_y}
    =
    \frac{1}{\sqrt{L_xL_y}}
    \sum_\mathbf{i}
    e^{-\im \mathbf{q\cdot i}}
    a_\mathbf{i}, 
\end{align}
with quasi-momenta $q_x=0,2\pi/L_x,...,2\pi(L_x-1)/L_x$ and $q_y=0,2\pi/L_y,...,2\pi(L_y-1)/L_y$, the Hamiltonian \eqref{eq:H} is diagonalized as 
\begin{align}
    H
    =
    \sum_{\bf q} 
    \epsilon_\mathbf{q}
    \tilde a_\mathbf{q}^\dag 
 \tilde a_\mathbf{q},
\end{align}
where the single-particle dispersion is $\epsilon_\mathbf{q}=-\cos q_x-\cos q_y$.
\par
\begin{figure}
    \centering
    \includegraphics[width=0.4\textwidth]{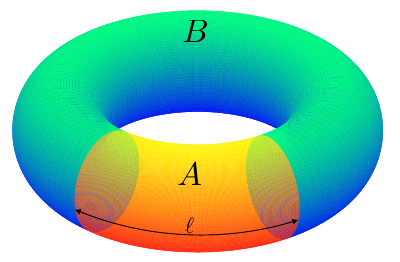}
    \caption{Schematic representation of the two-dimensional system that we study and of the subsystem $A$ considered.}
    \label{fig:subsystem}
\end{figure}

 
We consider the quantum quench described by the time-evolved state $|\psi(t)\rangle=e^{-\im t H}|\psi_0\rangle$ with an initial configuration $\ket{\psi_0}$ that is not an eigenstate of the Hamiltonian~\eqref{eq:H}. We take as a subsystem $A$ a periodic strip of length $\ell$, as depicted in Fig.~\ref{fig:subsystem}. That is, subsystem $A$ is the set of sites $\mathbf{i}$ satisfying $i_x\in[0, \ell-1]$. The state of $A$ is  described by the reduced density matrix 
\begin{align}
    \rho_A(t) 
    = 
    \Tr_B\left(|\psi(t)\rangle
    \langle\psi(t)|\right),
\end{align}
where $\Tr_{B}$ denotes the trace over the subsystem $B$. 
The R\'enyi entanglement entropy, 
\begin{align}
    S_n(\rho_A)=\frac{1}{1-n}\log\Tr(\rho_A^n),
    \label{eq:RE_def}
\end{align}
measures the degree of entanglement between subsystems $A$ and $B$. In the limit $n\to 1$, it gives the von Neumann entanglement entropy,
\begin{align}
S_1(\rho_A)=\lim_{n\to1}S_n(\rho_A)=-\Tr(\rho_A\log \rho_A).
\end{align}
Hereafter, we write $S_n(\rho_A)$ as $S_n$ unless explicitly stated. 

\par
In this paper, we consider  initial states $|\psi_0\rangle$ that satisfy Wick theorem. 
Therefore, since the post-quench Hamiltonian is a quadratic fermionic operator, the time-evolved reduced density matrix $\rho_A(t)$ is Gaussian and it is fully characterized by the two-point correlation matrix restricted to subsystem $A$~\cite{peschel-02},
\begin{align}
    \Gamma_\mathbf{i,i'}(t)
    &=
    2
    \bra{\psi(t)}\mathbf{a}_\mathbf{i}^\dag
    \mathbf{a}_{\bf i'} \ket{\psi(t)}
    -\delta_{\mathbf{i,i'}}, 
    \label{def:Gamma}
\end{align}
where $\mathbf{a}_\mathbf{i} = (a_\mathbf{i}^\dag,a_\mathbf{i})$ and $\mathbf{i,i'}\in A$. $\Gamma$ is
a matrix of dimension $2V_A\times 2V_A$, and $V_A$ is the size of the subsystem $A$, $V_A=\ell L_y$.
Using the standard algebra of Gaussian operators, the entanglement entropy can be expressed in terms of the two-point correlation matrix $\Gamma$ as \cite{ep-08}
\begin{align}
    S_n 
    = 
    \frac{1}{2(1-n)}
    \Tr
    \log\qty[
    \qty(\frac{I+\Gamma}{2})^n 
    +\qty(\frac{I-\Gamma}{2})^n 
    ], 
    \label{eq:S_n=tr_ln_Gamma}
\end{align}
where $I$ is the $2V_A\times 2V_A$ identity matrix. For finite $V_A$, we can compute the trace in Eq.~\eqref{eq:S_n=tr_ln_Gamma} and obtain the exact value of the entanglement entropy by numerically diagonalizing the two-point correlation matrix $\Gamma$. 
We will use this method as a benchmark of the analytic results obtained in the following sections. 
\par 
To perform analytical calculations, it is useful to write the right-hand side of Eq.\,\eqref{eq:S_n=tr_ln_Gamma} as a Taylor series in the moments $\Tr[\Gamma^m]$. To this end, we introduce the function $h_n(x)$,
\begin{equation}\label{eq:h_n}
    h_n(x) 
    = 
    \frac{1}{1-n}
    \log\qty[
    \qty(\frac{1+x}{2})^n
    +
    \qty(\frac{1-x}{2})^n
    ].
\end{equation}   
This function can be expanded as
\begin{equation}
  h_n(x)=\sum_{m=0}^\infty a_n(2m) x^{2m},
\end{equation}
and, therefore, Eq.~\eqref{eq:S_n=tr_ln_Gamma} can be rewritten in the form 
\begin{align}
    S_n 
    = 
    \frac{1}{2}
    \sum_{m=0}^{\infty}
    a_n(2m) \Tr(\Gamma^{2m}).
    \label{eq:S_n_expand_Gamma}
\end{align}
As we will see in the following sections, the precise form of the coefficients $a_n(2m)$ is never needed and hence will not be reported.  

\section{Dimensional reduction}
\label{sec:dimension_reduction}

In this section, we present the dimensional reduction approach that
we will employ in Sec.~\ref{sec:examples} to calculate analytically the time evolution of the R\'enyi entanglement entropy in different quantum quenches. 
The treatment is generically valid for initial states $\ket{\psi_0}$ that are invariant under $k$-site translations in the $y$-direction, with $k$ being a factor of $L_y$. 
For clarity, we present first the case $k=1$ and after we generalize straightforwardly to arbitrary $k$.

\subsection{One-site shift-invariant states in the transverse direction}

We start considering the case when the initial state $\ket{\psi_0}$ is translationally invariant in the $y$-direction. 
Since the Hamiltonian~\eqref{eq:H} preserves the translational symmetry, the time-evolved state $\ket{\psi(t)}$ is also invariant.

Given the geometry of the subsystem $A$ considered, it is useful to take the Fourier transform only along the $y$-direction by introducing the fermionic operators in a mixed space-momentum basis
\begin{align}
    c_{i_x,q_y}
    = 
    \frac{1}{\sqrt{L_y}}
    \sum_{i_y}
    e^{-\im q_y i_y}
    a_{i_x,i_y},
    \label{def:c_ix_qy}
\end{align}
which is the core of the dimensional reduction method. 
The two-point correlation function $\Gamma$ can be written as
\begin{equation}
\label{eq:Gamma_1_1}
\Gamma_{(i_x, j_y), (i_x', j_y')}
=\frac1{L_y}
\sum_{n_y=0}^{L_y-1} e^{{\rm i} \frac{2 \pi n_y}{L_y}(j_y-j_y')}(\Gamma_{q_y})_{i_x, i_x'},
\end{equation}
where $\Gamma_{q_y}$ is the $2\ell\times 2\ell$ two-point correlation matrix in the mixed space-momentum representation and the sum of $q_y=2\pi n_y/L_y$ runs on the $L_y$ allowed transverse modes.
As a consequence, modulo the Fourier transform which is a unitary operation, we have the decomposition
\begin{equation}
\Gamma\simeq\bigoplus_{n_y=0}^{L_y-1}\Gamma_{q_y},
\end{equation}
and so the entanglement entropy admits the decomposition in $L_y$ independent terms
\begin{gather}
S_n = \sum_{n_y=0}^{L_y-1}
    S_n(\Gamma_{q_y}),
    \label{eq:S_n_1-site}
\end{gather}
where we denoted by $S_n(\Gamma_{q_y})$ the R\'enyi entropy of the Gaussian state of a 1D chain with correlation matrix $\Gamma_{q_y}$.
Notice that the decomposition 
\eqref{eq:S_n_1-site} is valid for arbitrary values of $L_x$ and $L_y$, not necessarily in the thermodynamic limit.

\subsection{$k$-site shift-invariant states in the transverse direction}

Let us now assume that the initial state $\ket{\psi_0}$ is invariant under $k$-site translations in the $y$-direction with $k$ being a factor of $L_y$. 
Once again, since the Hamiltonian~\eqref{eq:H} preserves the translational symmetry, the time-evolved state $\ket{\psi(t)}$ is also invariant under $k$-site translations. 
This property is inherited by the two-point correlation matrix $\Gamma(t)$~\eqref{def:Gamma}, and its entries satisfy 
\begin{equation}\label{eq:trans_corr}
\Gamma_{(i_x,i_y+mk),(i_x',i_y'+mk)} = \Gamma_{(i_x,i_y),(i_x',i_y')}, \quad \forall m\in \mathbb{Z}.
\end{equation}
Therefore, it is convenient to decompose the index in the $y$-direction as $i_y=kj_y+p$, with $j_y=0,\dots, L_y/k-1$ and $p=0, \dots, k-1$, and then rearrange the entries of $\Gamma(t)$ in $k\times k$ blocks of the form
\begin{equation}\label{eq:corr_mat_rearranged}
\Gamma_{(i_x, j_y), (i_x', j_y')}(t)=2\bra{\psi(t)} \vec{\mathbf{a}}_{i_x, j_y}^\dagger \vec{\mathbf{a}}_{i_x', j_y'}\ket{\psi(t)}\\-\delta_{i_x, i_x'}\delta_{j_y, j_y'}
\end{equation}
where $\vec{\mathbf{a}}_{i_x, j_y}=(\mathbf{a}_{i_x, kj_y}, \mathbf{a}_{i_x,  kj_y+1},\dots, \mathbf{a}_{i_x, kj_y+k-1})$, with $i_x\in A$.

Once again, we take the Fourier transform only along the $y$-direction using Eq. \eqref{def:c_ix_qy}.  
The two-point correlation function $\Gamma$ can be then written as
\begin{equation}
\label{eq:Gamma_1}
\Gamma_{(i_x, j_y), (i_x', j_y')} =\frac{k}{L_y}
\sum_{n_y=0}^{\frac{L_y}k-1} 
e^{{\rm i} \frac{2 \pi k n_y}{L_y}(j_y-j_y')}(\mathcal{G}_{q_y}^{(k)})_{i_x, i_x'},
\end{equation}
where 
\begin{equation}\label{eq:mathcalG_Gamma_q_rel}
    (\mathcal{G}_{q_y}^{(k)})_{i_x,i_x'}=U (\Gamma_{q_y}^{(k)})_{i_x, i_x'}U^\dagger.
\end{equation}
Here $U$ is a unitary matrix with entries $U_{pp'}=e^{{\rm i}2\pi (\frac{p}{L_y}+\frac{pp'}{k})}/\sqrt{k}$ and $(\Gamma_{q_y}^{(k)})_{i_x, i_x'}$
is the $2k\ell\times 2k\ell$ two-point correlation matrix in the mixed space-momentum representation whose entries are rearranged in $k\times k$ blocks as
\begin{equation}
    (\Gamma_{q_y}^{(k)})_{i_x, i_x'}
    =
    2\bra{\psi(t)}
    \Vec{\mathbf{c}}_{i_x,q_y}^\dag 
    \Vec{\mathbf{c}}_{i_x',q_y}
    \ket{\psi(t)}
    -\delta_{i_x,i_x'},
    \label{eq:Gamma_ixqy_ixqy}
\end{equation}
where $q_y=\frac{2\pi n_y}{L_y}$ and
\begin{align}
    \Vec{\mathbf{c}}_{i_x,q_y}
    =
    (\mathbf{c}_{i_x,q_y}\,
    \mathbf{c}_{i_x,q_y+\frac{2\pi}{k}}\cdots
    \mathbf{c}_{i_x,q_y+\frac{2\pi(k-1)}{k}}).
    \label{eq:vec_c}
\end{align}
with $\mathbf{c}_{i_x,q_y}=(c_{i_x,q_y}^\dag,c_{i_x,-q_y})$. 

Given Eq.~\eqref{eq:Gamma_1}, the matrix $\Gamma$ takes the form
\begin{equation}
\Gamma=\frac{k}{L_y}\sum_{n_y=0}^{\frac{L_y}{k}-1}\mathcal{G}_{q_y}^{(k)}\otimes T_{q_y},
\end{equation}
where we have introduced the matrices $(T_{q_y})_{j_yj_y'}=e^{{\rm i}kq_y(j_y-j_y')}$. These matrices mutually commute and can be
diagonalized simultaneously by
\begin{equation}
V_{j_y, j_y'}=\frac{1}{\sqrt{k}}e^{{\rm i}2\pi j_yj_y'/k}
\end{equation}
such that $(VT_{q_y}V^\dagger)_{j_y, j_y'}=k \delta_{q_y, j_y}\delta_{q_y, j_y'}$. 
As a consequence, we have the decomposition
\begin{equation}\label{eq:Gamma_sum_sectors}
(I\otimes V)\Gamma(I\otimes V^\dagger)=\bigoplus_{n_y=0}^{\frac{L_y}{k}-1}\mathcal{G}_{q_y}^{(k)},
\end{equation}
showing that the two-point correlation matrix $\Gamma$ is block diagonal in the $q_y$ transverse momentum sectors. 
Plugging Eq.\,\eqref{eq:Gamma_sum_sectors} into Eq.\,\eqref{eq:S_n=tr_ln_Gamma}, and taking into account that $\mathcal{G}_{q_y}^{(k)}$ and $(\Gamma_{q_y}^{(k)})$ are related by a unitary transformation~\eqref{eq:mathcalG_Gamma_q_rel}, we finally obtain 
\begin{gather}
    S_n= 
    \sum_{n_y=0}^{\frac{L_y}{k}-1}
     S_n(\Gamma_{q_y}^{(k)})
      = 
    \frac{1}{2} \sum_{n_y=0}^{\frac{L_y}{k}-1}
    \sum_{m=0}^{\infty}
    a_n(2m) \Tr[(\Gamma_{q_y}^{(k)})^{2m}].
    \label{eq:S_n_k-site}
\end{gather}
Accordingly, the entanglement entropy in our 2D system is the sum of the single-interval entanglement entropies of $L_y/k$ one-dimensional fermionic chains, each univocally characterized by the correlation matrices $\Gamma_{q_y}^{(k)}$.

\section{Examples}
\label{sec:examples}

In this section, using the dimensional reduction approach described in Sec.~\ref{sec:dimension_reduction} and invoking results for 1D systems, we calculate the entanglement entropy in quantum quenches  from several initial states. 
In particular, we analytically derive its exact behavior in the ballistic regime in which $t,\ell\rightarrow\infty$ with $t/\ell$ fixed, taking the thermodynamic limit in the longitudinal direction, $L_x\to\infty$, with the transverse one $L_y$ finite. Here, we only present the results of the calculations, while their physical interpretation will be discussed in Sec.~\ref{sec:quasiparticle_picture}. 

For the concrete initial configurations that we will consider, the time evolution of the entanglement entropy can be calculated using Eq.~\eqref{eq:S_n_k-site}. The latter requires the knowledge of the matrix $\Gamma_{q_y}^{(k)}$ defined in Eq. \eqref{eq:Gamma_ixqy_ixqy}, which involves the mixed space-momentum
correlations $\langle c_{i_x, q_y}^\dagger c_{i_x', q_y'}\rangle$ and $\langle c_{i_x, q_y} c_{i_x', q_y'}\rangle$. Their time 
evolution can be easily computed employing the Heisenberg picture since the momentum modes $\tilde a_{\bf q}$ that diagonalize the post-quench Hamiltonian~\eqref{eq:H} evolve trivially in time as $\tilde a_{\bf q}(t)=e^{-{\rm i}t\epsilon_{\bf q}} \tilde a_{\bf q}$. Therefore, taking the partial Fourier transform in the $x$ direction, we have
\begin{multline}
\bra{\psi(t)}c_{i_x, q_y}^\dagger c_{i_x', q_y'}\ket{\psi(t)}=\\
\frac{1}{L_x}\sum_{q_x, q_x'}
e^{-{\rm i}(q_xi_x-q_x' i_x')}
e^{{\rm i} t(\epsilon_{\bf q}-\epsilon_{\bf q'})}\bra{\psi_0} \tilde a_{\bf q}^\dagger \tilde a_{{\bf q}'}\ket{\psi_0}
\end{multline}
and 
\begin{multline}
\bra{\psi(t)}c_{i_x, q_y} c_{i_x', q_y'}\ket{\psi(t)}=\\
\frac{1}{L_x}
\sum_{q_x, q_x'}e^{{\rm i}(q_xi_x+q_x' i_x')} e^{-{\rm i}t(\epsilon_{\bf q}+\epsilon_{\bf q'})} 
\bra{\psi_0} \tilde a_{{\bf q}} \tilde a_{{\bf q}'}\ket{\psi_0}.
\end{multline}
Writing now the operators $\tilde a_{\bf q}$ and $\tilde a_{\bf q}^\dagger$ in terms of the real space ones, $a_{\bf i}$ and $a_{\bf i}^\dagger$, we
find
\begin{multline}\label{eq:cdaggerc_corr}
\bra{\psi(t)} c_{i_x, q_y}^\dagger c_{i_x', q_y'}\ket{\psi(t)}=
\frac{1}{L_x^2 L_y}\sum_{q_x, q_x'} \sum_{{\bf j}, {\bf j}'}
e^{-{\rm i}(q_x i_x - q_x' i_x')} 
\\
\times e^{{\rm i}t(\epsilon_{\bf q}-\epsilon_{{\bf q}'})}e^{{\rm i}({\bf q}\cdot {\bf j}-{\bf q}'\cdot {\bf j}')}\bra{\psi_0} a_{\bf j}^\dagger a_{{\bf j}'}\ket{\psi_0}
\end{multline}
and
\begin{multline}\label{eq:cc_corr}
\bra{\psi(t)} c_{i_x, q_y} c_{i_x', q_y'}\ket{\psi(t)}= 
\frac{1}{L_x^2 L_y}\sum_{q_x, q_x'}\sum_{{\bf j}, {\bf j}'}
e^{{\rm i} (q_x i_x + q_x' i_x')} 
\\
\times e^{-{\rm i}t(\epsilon_{\bf q}+\epsilon_{{\bf q}'})}e^{-{\rm i}({\bf q}\cdot {\bf j}-{\bf q}'\cdot {\bf j}')}\bra{\psi_0} a_{\bf j} a_{{\bf j}'}\ket{\psi_0}.
\end{multline}
Eqs.~\eqref{eq:cdaggerc_corr} and \eqref{eq:cc_corr} recast the time evolution of the correlators $\langle c_{i_x, q_y}^\dagger c_{i_x', q_y'}\rangle$, $\langle c_{i_x, q_y} c_{i_x', q_y'}\rangle$ and, consequently of the matrix $\Gamma_{q_y}^{(k)}$, in terms of the $t=0$ correlators $\langle a_{\bf j}^\dagger a_{{\bf j}'}\rangle$ and $\langle a_{\bf j} a_{{\bf j}'}\rangle$.

\subsection{Collinear Mott insulator state}
\label{subsec:CAF}

\begin{figure}[t]
\includegraphics[width=0.48\textwidth]{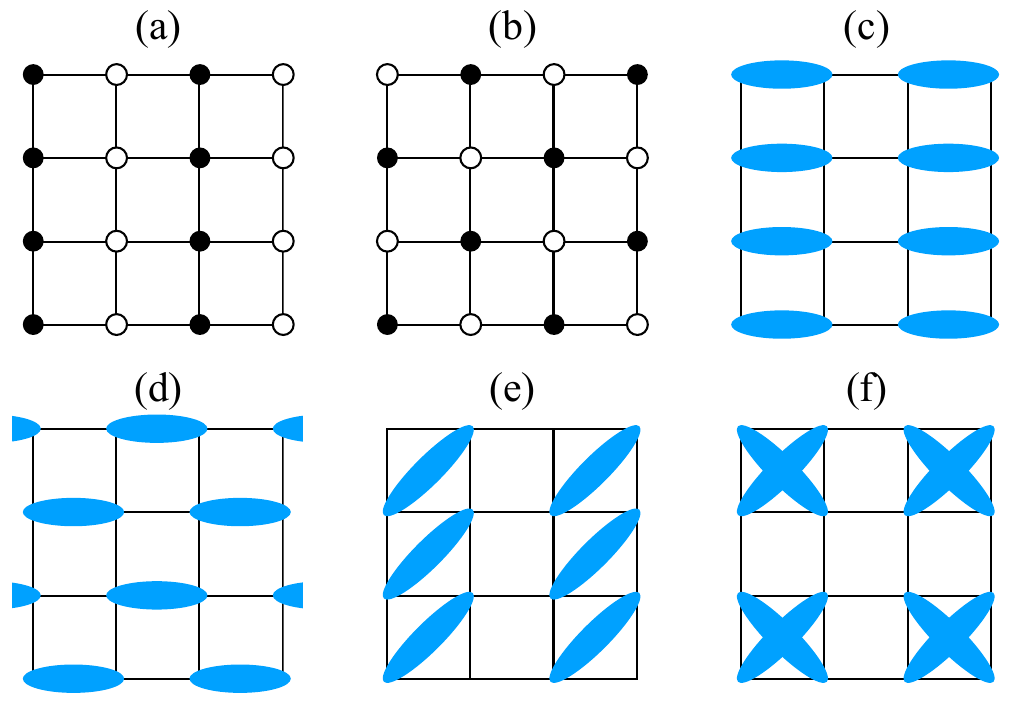}
\caption{Schematic representations of the states that we consider as initial configurations for the quantum quenches. They are: (a) the collinear Mott state, (b) the Mott state, (c) the collinear dimer state, (d) the staggered dimer state, (e) the diagonal dimer state, and (f) the crossed dimer state. The black and white dots represent the occupied and empty sites respectively. The blue ellipses represent singlet pairs.}
    \label{fig:initial_states}
\end{figure}

We start with the quantum quench from the collinear Mott insulator state, which is defined as 
\begin{align}
    \ket{\mathrm{CM}}
    =
    \prod_{i_x=0}^{\frac{L_x}{2}-1}
    \prod_{i_y=0}^{L_y-1}
    a_{2i_x,i_y}^\dag
    \ket{0},
    \label{eq:|CAF>}
\end{align}
where $\ket{0}$ is the space vacuum state; i.e., $a_{\bf i}\ket{0}=0$ for all ${\bf i}$. %
We schematically represent this configuration in Fig.\,\ref{fig:initial_states}~(a). 
The collinear Mott insulator state \eqref{eq:|CAF>} is a product state and, therefore, the entanglement entropy for any bipartition is zero. 
 
According to Eq.~\eqref{eq:S_n_k-site}, since the state \eqref{eq:|CAF>} is invariant under single-site translations in the $y$-direction, the entanglement entropy after the quench can be calculated from the correlation matrix $\Gamma_{q_y}^{(1)}$. The 
entries of this matrix, see Eq.~\eqref{eq:Gamma_ixqy_ixqy}, are given by the correlators $\langle c_{i_x, q_{y}}^\dagger c_{i_x', q_y'}\rangle$ and $\langle c_{i_x, q_{y}} c_{i_x', q_y'}\rangle$, whose time evolution can be obtained with Eqs.~\eqref{eq:cdaggerc_corr} and \eqref{eq:cc_corr} in terms of $\langle a_{\bf j}^\dagger a_{\bf j'}\rangle$  $\langle a_{\bf j} a_{\bf j'}\rangle$ for the initial state, which in this  case read
\begin{equation}
\label{eq:ada_CM}
    \bra{\rm CM} 
    a_{\bf j}^\dag a_{\bf j'}
    \ket{\rm CM}=
    \frac{\delta_{\bf j,j'}}{2}
    [1+(-1)^{j_x}],
\end{equation}
and
\begin{equation}\label{eq:aa_CM}
    \bra{\rm CM} 
    a_{\bf j} a_{\bf j'}
    \ket{\rm CM}
    =0. 
\end{equation}
Note that the pairing correlation functions such as $\langle c_{i_x,q_y} c_{i_x',-q_y} \rangle$ vanish because the state \eqref{eq:|CAF>} has a definite number of excitations.  

Plugging Eqs.~\eqref{eq:ada_CM} and \eqref{eq:aa_CM} in Eqs.~\eqref{eq:cdaggerc_corr} and \eqref{eq:cc_corr} respectively, 
we obtain
\begin{gather}
    (\Gamma_{q_y}^{(1)})_{i_x,i_x'}(t)
    = 
    \mqty(
    -C_{i_x',i_x}(t)&0\\0& C_{i_x,i_x'}(t)
    ),
    \label{eq:Gamma=C}
\end{gather}
where $C_{i_x,i_x'}(t)$ is the 1D correlation matrix after the quench to the tight binding fermionic chain from the N\'eel state, see, e.g., \cite{parez2021}. In the thermodynamic limit $L_x\to\infty$, it reads
\begin{gather}
    C_{i_x,i_x'}(t)
    = 
    (-1)^{i_x'}
    \int_0^{2\pi}
    \frac{\dd q_x}{2\pi}
    e^{-\im q_x(i_x-i_x')-2\im t \cos q_x}.
    \label{eq:C_qy_Neel}
\end{gather}

\begin{figure}[t]
\includegraphics[width=0.45\textwidth]{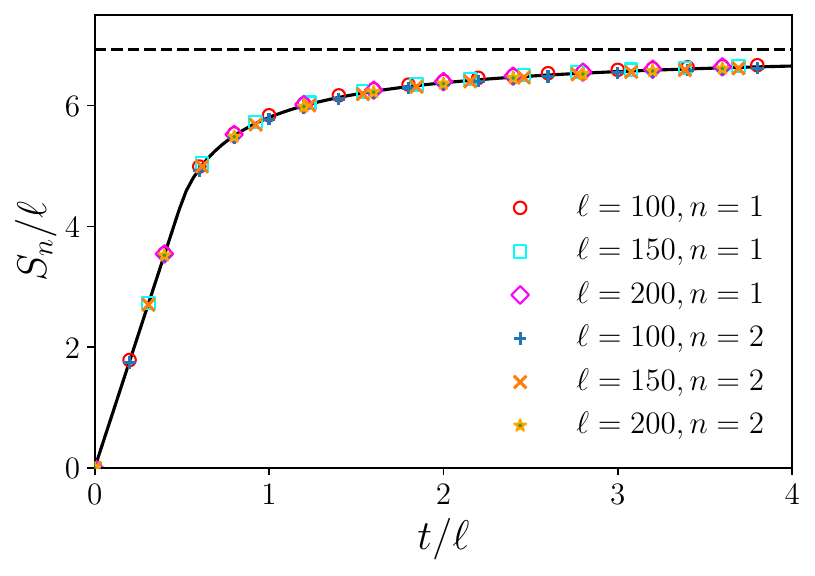}
    \caption{Time evolution of the R\'enyi entanglement entropy in the quantum quench from the collinear Mott insulator and Mott insulator states. 
    The solid line is the analytic result in Eq.\,\eqref{eq:S_n_2D_Neel_sp_limit}.
    The symbols are the exact value of the entropy obtained numerically using Eq.~\eqref{eq:S_n=tr_ln_Gamma}. We take $L_y=10$.}
    \label{fig:S_1_2D_Neel}
\end{figure}

Plugging Eq.\,\eqref{eq:Gamma=C} into Eq.\,\eqref{eq:S_n_k-site} with $k=1$, we obtain 
\begin{align}
    S_{n} = 
    \sum_{n_y=0}^{L_y-1}
S_n(C)= L_y S_n(C).
    \label{eq:S_n_Neel_2}
\end{align}
The asymptotic form of the 1D entanglement entropy with correlation matrix \eqref{eq:C_qy_Neel} in the ballistic regime is known~\cite{parez2021},
\begin{align}
   S_n(C)
    \simeq \log (2)
    \int_0^{2\pi}
    \frac{\dd q_x}{2\pi}
    \min(\ell,2t|v_x(q_x)|).
\end{align} 
where $v_x(q_x)=\partial_{q_x}\epsilon_\mathbf{q}=\sin q_x$ is the fermion velocity in the $x$-direction and, by $\simeq$, we always mean equal in the thermodynamic and ballistic limits. 
Hence for the 2D model we obtain that
\begin{align}
    S_n(t)
    \simeq  
    L_y \log (2)
    \int_0^{2\pi}
    \frac{\dd q_x}{2\pi}
    \min(\ell,2t|v_x(q_x)|).
    \label{eq:S_n_2D_Neel_sp_limit}
\end{align}
The expression above shows that the entanglement entropy linearly increases in time for $t<\ell/\max(2v_x)=\ell/2$, while for $t\gg \ell/2$ it saturates to a constant value, 
\begin{align}
    \lim_{t\rightarrow\infty}
    S_n
    \simeq 
    V_A \log(2).
    \label{eq:S_n_saturate_2D_Neel}
\end{align}
In Fig.~\ref{fig:S_1_2D_Neel}, we report the time evolution of the entanglement entropy for the quench from the collinear Mott insulator state. 
The curve is the analytic result~\eqref{eq:S_n_2D_Neel_sp_limit}, which agrees well with the exact numerical  data obtained using Eq.~\eqref{eq:S_n=tr_ln_Gamma}. 

\subsection{Mott insulator state} 
\label{subsection:NAF}

We now consider the quantum quench from the Mott insulator state, which is defined as 
\begin{align}
    \ket{\mathrm{M}}
    =
\prod_{i_x+i_y=\mathrm{even}}
    a_\mathbf{i}^\dag \ket{0}.
    \label{eq:|NAF>}
\end{align}
This state is represented schematically in Fig.~\ref{fig:initial_states} (b). 
As in the case of the collinear Mott insulator state discussed in Sec.\,\ref{subsec:CAF}, this configuration is also a product state and, therefore, the entanglement entropy at $t=0$ is zero.
\par 
Since the Mott insulator state is invariant under two-site translations in the $y$-direction, the entanglement entropy after the quench can be obtained by applying Eq.~\eqref{eq:S_n_k-site} once we have determined the time evolution of the correlation matrix $\Gamma_{q_y}^{(2)}$. Using Eqs.~\eqref{eq:cdaggerc_corr} and \eqref{eq:cc_corr}, we only need the two-point spatial correlations in the initial configuration to calculate it. For the Mott insulator state, 
\begin{align}
    \bra{\rm M}
    a_{\bf j}^\dag a_{\bf j'}
    \ket{\rm M}
    =
    \frac{\delta_{\bf j,j'}}{2}
    [1+(-1)^{j_x+j_y}],
\end{align}
and
\begin{align}
    \bra{\rm M}
    a_{\bf j} a_{\bf j'}
    \ket{\rm M}
    =0.
    \label{eq:<cc>_Neel}
\end{align}
If we insert them in Eqs.~\eqref{eq:cdaggerc_corr} and \eqref{eq:cc_corr} respectively, we obtain that the matrix $\Gamma_{q_y}^{(2)}$ is
\begin{multline}
(\Gamma_{q_y}^{(2)})_{i_x,i_x'}(t)
    =
    U
    e^{\im t \cos q_y \sigma_z\otimes \sigma_z}
    \\
    \times 
    \mqty(-C_{i_x',i_x}(t)&0\\ 0& C_{i_x,i_x'}(t))
    \otimes \sigma_x 
    e^{-\im t \cos q_y \sigma_z \otimes \sigma_z}
    U^{-1},
\end{multline}
where the matrix $C$ is the same as in Eq.~\eqref{eq:Gamma=C}, $\sigma_\mu$ are the Pauli matrices, and $U=I/2+\sum_{\mu=x,y,z} \sigma_\mu\otimes \sigma_\mu/2$. 
In the thermodynamic limit $L_x\to \infty$, $C$ is given by  Eq.\,\eqref{eq:C_qy_Neel}.
Plugging this result into Eq.\,\eqref{eq:S_n_k-site} with $k=2$ and using $\Tr \sigma_x^{2m}=2$, we obtain 
\begin{align}
    S_n    &=L_y S_n(C), 
    \label{eq:S_n_AF}
\end{align}
which coincides with the expression found in 
Eq.~\eqref{eq:S_n_Neel_2} for the collinear 
Mott insulator state \eqref{eq:|CAF>}. 
The matrix $C$ is the same in the Mott insulator and in the collinear Mott insulator states, both at finite size $L_x$ and when we take $L_x\to \infty$, and therefore the entanglement entropy presents exactly the same time evolution in both quenches and for this reason we do not report any numerical test.

\subsection{Collinear dimer state}

Let us now analyze the quantum quench from the collinear dimer state,
\begin{align}
    \ket{\mathrm{CD}}
    =
    \prod_{i_x=0}^{\frac{L_x}{2}-1}
    \prod_{i_y=0}^{L_y-1}
    \frac{a_{2i_x,i_y}^\dag-a_{2i_x+1,i_y}^\dag}{\sqrt{2}}
    \ket{0},
    \label{eq:|CV>}
\end{align}
which is represented in Fig.\,\ref{fig:initial_states} (c). 
Unlike the previous examples, this configuration is not a product state for each site, but the $(2i_x,i_y)$-th and $(2i_x+1,i_y)$-th pairs of sites are entangled by the singlet pairing. However, since these singlet pairs do not cross the boundaries of the subsystem considered (because we choose $\ell$ to be even and its endpoints to be also endpoints of singlets), we have that the entanglement
entropy is zero before the quench.
\par 
The collinear dimer state \eqref{eq:|CV>} has one-site translational symmetry in the $y$-direction. Since it is invariant under two-site translations in the $x$-direction, it is convenient to rearrange the entries of the correlation matrix $\Gamma_{q_y}^{(1)}$, which enters in the computation of the entropy~\eqref{eq:S_n_k-site}, as
\begin{multline}   (\Gamma_{q_y}^{(1)})_{i_x,i_x'}
    =
    2\Big\langle
    \mqty(
    \mathbf{c}_{2i_x,q_y}^\dag
    \\
    \mathbf{c}_{2i_x+1,q_y}^\dag
    )
    (
    \mathbf{c}_{2i_x',q_y}
    \,
    \mathbf{c}_{2i_x'+1,q_y}
    )
    \Big\rangle 
    \\
    \quad-\delta_{i_x,i_x'}I_4,
    \label{eq:Gamma_qy_CV}
\end{multline}
with $i_x,i_x'\in [0,\ell/2-1]$. The correlators in $\Gamma_{q_y}^{(1)}$ can be calculated using Eqs.\,\eqref{eq:cdaggerc_corr} and \eqref{eq:cc_corr} with 
\begin{multline}
    \bra{\rm CD}a_{\bf j}^\dag a_{\bf j'}
    \ket{\rm CD}
    =
    \frac{1}{2}\delta_{\bf j,j'}
    -\frac{1}{4}\delta_{j_y,j_y'}
    \delta_{j_x\pm1,j_x'}
    \nonumber\\
    \pm 
    \frac{(-1)^{j_x}}{4}
    \delta_{j_y,j_y'}
    \delta_{j_x\pm1,j_x'},
\end{multline}
and
\begin{equation}
    \bra{\rm CD}a_{\bf j} a_{\bf j'}
    \ket{\rm CD}
    =0.
\end{equation}
In this way, once we have properly organized all the entries of $\Gamma_{q_y}^{(1)}$, we find that it takes the form
\begin{align}   (\Gamma_{q_y}^{(1)})_{i_x,i_x'}
    =
    U
    \mqty(-T_{i_x',i_x}^{\rm D}&0\\0&T_{i_x,i_x'}^{\rm D})
    U^\dag,    
    \label{eq:Gamma=UTU}
\end{align}
where $U=I_4/2+\sum_{\mu=x,y,z}\sigma_\mu\otimes\sigma_\mu/2$ is a $4\times4$ unitary matrix and $T^{\rm D}$ is equal to the 1D two-point correlation matrix of the quench to the tight binding fermionic chain from the dimer state (see e.g. Ref.~\cite{parez2021}). In the thermodynamic limit $L_x\to\infty$, $T^{\rm D}$ is a block Toeplitz matrix,
\begin{align}
    T_{i_x,i_x'}^{\rm D}
    =
    \int_0^{2\pi}
    \frac{\dd q_x}{2\pi}
    e^{-2\im q_x(i_x-i_x')}
    g_{\rm D}(q_x), 
    \label{eq:T_CV}
\end{align}
generated by the $2\times 2$ symbol $g_{\rm D}(q_x)$
\begin{align}
    g_{\rm D}(q_x)
    =
    e^{\im \sigma_z\frac{q_x}{2}}
    (\sigma_- 
    \sin q_x 
    e^{-2\im t \cos q_x}
    -\sigma_x
    \cos q_x
    )
    e^{-\im \sigma_z\frac{q_x}{2}}, 
    \label{eq:g_CV}
\end{align}
with $\sigma_\pm=\sigma_y\pm \im \sigma_z$. 
\par 
Plugging Eq.\,\eqref{eq:Gamma=UTU} into Eq.\,\eqref{eq:S_n_k-site}, we obtain 
\begin{align}
    S_n=
    L_y S_n(T^{\rm D}).
    \label{eq:S_CV}
\end{align}
In the ballistic regime, the asymptotic form of entropy $S_n(T^{\rm D})$ is known and reads~\cite{parez2021}
\begin{equation}
    S_n(T^{\rm D})
    \simeq 
    \int_0^{2\pi}
    \frac{\dd q_x}{2\pi}
    h_n(\cos q_x) \min(\ell,2t|v_x(q_x)|),
    \label{eq:trC_qy^2m_VBS}
\end{equation} 
providing that the R\'enyi entanglement entropy after the quench in 2D behaves 
as
\begin{align}
    S_{n}
    \simeq  
    L_y
    \int_0^{2\pi}
    \frac{\dd q_x}{2\pi}
    h_n(\cos q_x) \min(\ell,2t|v_x(q_x)|). 
    \label{eq:S1_sp_limit_vbs}
\end{align}
Therefore, the entanglement entropy increases linearly in time for $t<\ell/2$, while it approaches the constant value 
\begin{align}
    \lim_{t\rightarrow\infty}
    S_{n}
    \simeq 
    V_A \int_0^{2\pi} 
    \frac{\dd q_x}{2\pi}
    h_n(\cos q_x)
    \label{eq:S1_saturation_vbs}
\end{align}
at large times $t\gg \ell/2$. 
\par
\begin{figure}
    \includegraphics[width=0.45\textwidth]{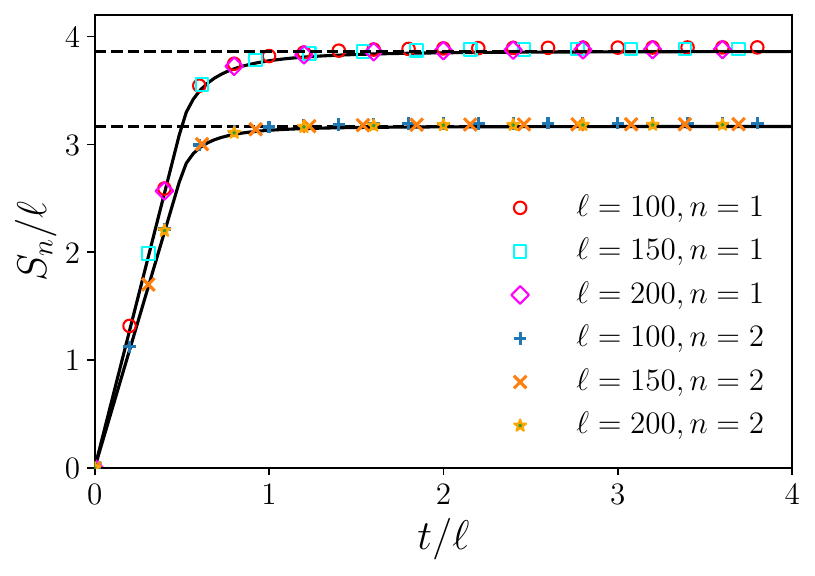}
    \caption{R\'enyi entanglement entropy as a function of $t/\ell$ in a quench from the collinear dimer state for $n\to 1$ and $n=2$ and different subsystem lengths $\ell$. 
    The solid lines correspond to the analytic prediction obtained in Eq.~\eqref{eq:S1_sp_limit_vbs}. The dashed lines indicate the saturation value~\eqref{eq:S1_saturation_vbs}.
    The symbols are the exact numerical value of the entropies calculated employing Eq.~\eqref{eq:S_n=tr_ln_Gamma}. We take $L_y=10$.}
    \label{fig:S1_vbs}
\end{figure}

 
In Fig.~\ref{fig:S1_vbs}, we check the validity of the analytic result~\eqref{eq:S1_sp_limit_vbs}. We plot it for $n\to 1$ and $n=2$, as a function of time (solid curves) and we compare with the exact numerical value computed using Eq.~\eqref{eq:S_n=tr_ln_Gamma}.

\subsection{Staggered dimer state}

We now investigate a modification of the previous initial configuration, the staggered dimer state, which is defined as 
\begin{align}
    \ket{\mathrm{SD}}
    =   \prod_{i_x+i_y=\mathrm{even}} 
    \frac{a_{i_x,i_y}^\dag-a_{i_x+1,i_y}^\dag}{\sqrt{2}}
    \ket{0}
    \label{eq:|SV>}
\end{align}
and illustrated in Fig.\,\ref{fig:initial_states} (d). 
In this case, there are $L_y$ singlet pairs crossing the boundary of the subsystem and, consequently, the initial entanglement entropy is $S_n=L_y\log(2)$ at $t=0$.
This initial offset is subleading (and hence negligible) in the ballistic limit because it does not scale with the volume $V_A=\ell L_y$. 
\par 
The staggered dimer state \eqref{eq:|SV>} is invariant under two-site translations in the $y$-direction, i.e. $k=2$. 
Since it is also invariant under two-site translations in the $x$-direction, it is convenient to rearrange the entries of the matrix $\Gamma_{q_y}^{(2)}$  as 
\begin{multline}
(\Gamma_{q_y}^{(2)})_{i_x,i_x'}
    =
    2\Big\langle
    \mqty(\Vec{\bf c}_{2i_x,q_y}^\dag
    \\
    \Vec{\bf c}_{2i_x+1,q_y}^\dag
    )
    (\Vec{\bf c}_{2i_x',q_y}
    \,
    \Vec{\bf c}_{2i_x'+1,q_y}
    )
    \Big\rangle 
    \\
    -\delta_{i_x,i_x'}I_8, 
    \label{eq:Gamma_qy_rearrange}
\end{multline}
with $i_x,i_x'\in[0,\ell/2-1]$. 
The definition of $\Vec{\bf c}_{i_x,q_y}$ is given in Eq.\,\eqref{eq:vec_c}. The entries of $\Gamma_{q_y}^{(2)}$ can be calculated by plugging the initial state correlators
\begin{multline}
    \bra{\rm SD}
    a_{\bf j}^\dag a_{\bf j'}
    \ket{\rm SD}
    =
    \frac{\delta_{\bf j,j'}}{2}
    -\frac{1}{4}
    \delta_{j_y,j_y'}
    \delta_{j_x\pm1,j_x'}
    \\
    \pm 
    \frac{(-1)^{j_x+j_y}}{4}
    \delta_{j_y,j_y'}
    \delta_{j_x\pm1,j_x},
\end{multline}
and $\bra{\rm SD}a_{\bf j} a_{\bf j'}\ket{\rm SD}=0$,
into Eqs.\,\eqref{eq:cdaggerc_corr} and \eqref{eq:cc_corr}. 
Then we find that $\Gamma_{q_y}^{(2)}$ is of the form
\begin{align}
(\Gamma_{q_y}^{(2)})_{i_x,i_x'}
    =
    U \mqty(-(T_{q_y+\pi}^{\rm SD})_{i_x',i_x}&0\\0& (T_{q_y}^{\rm SD})_{i_x,i_x'}) U^\dag, 
    \label{eq:Gamma_qy_VBS2}
\end{align}
where $U=I_8/2+\sum_{\mu=x,y,z}\sigma_\mu\otimes I_2\otimes \sigma_\mu/2$ is a $8\times 8$ unitary matrix and, in the thermodynamic limit $L_x\to\infty$, $T^{\rm SD}_{q_y}$ is a block Toeplitz matrix,
\begin{align}
    (T_{q_y})_{i_x,i_x'}^{\rm SD}
    =\int_0^{2\pi}
    \frac{\dd q_x}{2\pi}
    e^{-2\im  q_x(i_x-i_x')}
    g_{q_y}^{\rm SD}(q_x), 
    \label{eq:T_SV}
\end{align}
with $4\times 4$ symbol
\begin{multline}
    g_{q_y}^{\rm SD}(q_x)
    =
    -(e^{-\im \sigma_z t\cos q_y}\otimes e^{\im \sigma_z\frac{q_x}{2}})\\
    \quad (I\otimes \sigma_x\cos q_x+\sigma_x\otimes \sigma_+e^{-2\im t \cos q_x})
    \\
     (e^{\im \sigma_z t\cos q_y}\otimes e^{-\im \sigma_z\frac{q_x}{2}}).
    \label{eq:g_qy_VBS2}
\end{multline}
\par
\begin{figure}
\includegraphics[width=0.45\textwidth]{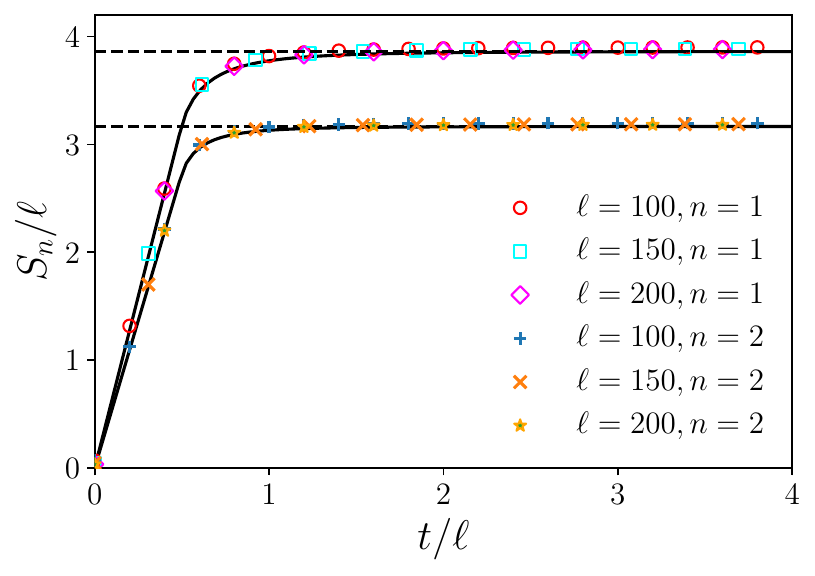}
    \caption{R\'enyi entanglement entropy as a function of $t/\ell$ in a quench starting from the staggered dimer state for $n\rightarrow1$ and $n=2$ and different subsystem lengths $\ell$. The solid lines correspond to the analytical prediction obtained in Eq.\,\eqref{eq:SD_RE_SP_limit}. The dashed lines indicate the saturation values \eqref{eq:S1_saturation_vbs}. The symbols are the exact numerical value of the entropies calculated employing Eq.~\eqref{eq:S_n=tr_ln_Gamma}. We take $L_y=10$.}
    \label{fig:SD_RE}
\end{figure}

 
In the ballistic limit, the asymptotic form of the moments $\Tr\,[ (\Gamma_{q_y}^{(2)})^{2m}]$ with Eq.\,\eqref{eq:Gamma_qy_VBS2} can be obtained by an analogous procedure used in Ref. ~\cite{parez2021} to derive Eq.\,\eqref{eq:trC_qy^2m_VBS}.
A tedious but straitforward calculation leads to the final result 
\begin{multline}
\Tr[(\Gamma_{q_y}^{(2)})^{2m}]
    \simeq 
    4\ell 
    -4\int_0^{2\pi} 
    \frac{\dd q_x}{2\pi}
    \\
    \times 
    [1-(\cos q_x)^{2m}]
    \min(\ell,2t|v_x(q_x)|). 
\end{multline}
Plugging it into Eq.\,\eqref{eq:S_n_k-site} with $k=2$, we arrive at 
\begin{align}
    S_n
    \simeq 
    L_y
    \int_0^{2\pi}
    \frac{\dd q_x}{2\pi}
    h_n(\cos q_x) 
    \min(\ell,2t|v_x(q_x)|). 
    \label{eq:SD_RE_SP_limit}
\end{align}
This expression is equal to Eq.\,\eqref{eq:S1_sp_limit_vbs} for the collinear dimer state. This coincidence only occurs in the limit
$L_x\to \infty$, for finite $L_x$ the entanglement entropy in these two quenches is different.
We check the validity of the analytical prediction \eqref{eq:SD_RE_SP_limit} in Fig.\,\ref{fig:SD_RE}. 
It shows that, for $n\rightarrow1$ and $n=2$, Eq.\,\eqref{eq:SD_RE_SP_limit} agrees well with the exact results obtained by evaluating numerically Eq.\,\eqref{eq:S_n=tr_ln_Gamma}.

\subsection{Diagonal dimer state} 
\label{subsec:diagonal-dimer}

We next take as initial configuration the diagonal dimer state,
\begin{align}
    \ket{\mathrm{DD}}
    = 
    \prod_{i_x=0}^{\frac{L_x}{2}-1}
    \prod_{i_y=0}^{L_y-1}
    \frac{a_{2i_x,i_y}^\dag-a_{2i_x+1,i_y+1}^\dag}{\sqrt{2}}
    \ket{0}.  
    \label{eq:|psi(0)>_DD}
\end{align}
An illustration of it can be found in Fig.\,\ref{fig:initial_states} (e). 

The diagonal dimer state is invariant under one-site translations in the $y$-direction and under two-site translations in the $x$-direction. Therefore, we rearrange the entries of $\Gamma_{q_y}^{(1)}$ as we have done in Eq.\,\eqref{eq:Gamma_qy_CV}.
The two-point spatial correlations in the initial state are in this
case,
\begin{multline}
    \bra{\rm DD}
    a_{\bf j}^\dag a_\mathbf{j'}
    \ket{\rm DD}
    =
    \frac{\delta_{\bf j,j'}}{2}
    -\frac{1}{4}\delta_{j_x\pm1,j_x'}
    \delta_{j_y\pm1,j_y'}
    \\
    \mp 
    \frac{(-1)^{j_x}}{4}
    \delta_{j_x\pm1,j_x'}
    \delta_{j_y\pm1,j_y'},
\end{multline}
and $\bra{\rm DD}a_{\bf j} a_\mathbf{j'}\ket{\rm DD}=0$.
Inserting them in Eqs.~\eqref{eq:cdaggerc_corr} and \eqref{eq:cc_corr}, 
we find that the matrix $\Gamma_{q_y}^{(1)}$ can be written as
\begin{gather}
(\Gamma_{q_y}^{(1)})_{i_x,i_x'}
    = 
    U\mqty(-(T_{q_y}^{\rm DD})_{i_x',i_x}&0\\0&(T_{-q_y}^{\rm DD})_{i_x,i_x'})U^\dag,
    \label{eq:C_qy_D_Toeplitz}
\end{gather}
where $U=I_4/2-\sum_{\mu=x,y,z}\sigma_\mu\otimes\sigma_\mu/2$. In the thermodynamic limit $L_x\to\infty$, $T^{\rm DD}_{q_y}$ is the block Toeplitz matrix
\begin{align}
    (T_{q_y}^{\rm DD})_{i_x,i_x'}
    &=
    \int_0^{2\pi}
    \frac{\dd q_x}{2\pi}
    e^{-2\im q_x(i_x-i_x')}
    g_{q_y}^{\rm DD}(q_x),
    \label{eq:C_qy_D_Toeplitz_}
\end{align}
with symbol
\begin{multline}
    g_{\pm q_y}^{\rm DD}(q_x)
    = 
    -e^{\im \frac{q_x}{2}\sigma_z}
    [\sigma_x \cos (q_x+ q_y)
    \\
    +e^{-2\im t \cos q_x}
    \sigma_+ \sin(q_x+ q_y)]
    e^{-\im \frac{q_x}{2}\sigma_z}.
    \label{eq:g_qy(qx)_D}
\end{multline}
\par 
The calculation of the moments $\Tr\,[(\Gamma_{q_y}^{(1)})^{2m}] $ with Eq.\,\eqref{eq:C_qy_D_Toeplitz} in the ballistic limit
is analogous to the computation of Eq.\,\eqref{eq:trC_qy^2m_VBS}. We find
\begin{multline}
\Tr[(\Gamma_{q_y}^{(1)})^{2m}]
    \simeq 
    2\ell 
    -
    2\int_0^{2\pi}
    \frac{\dd q_x}{2\pi}
    \{1-[\cos (q_x+q_y)]^{2m}\}
    \\
    \times\min(\ell,2t|v_x(q_x)|).
    \label{eq:C_qy_asymptotic_D}
\end{multline}

\begin{figure}
\includegraphics[width=0.45\textwidth]{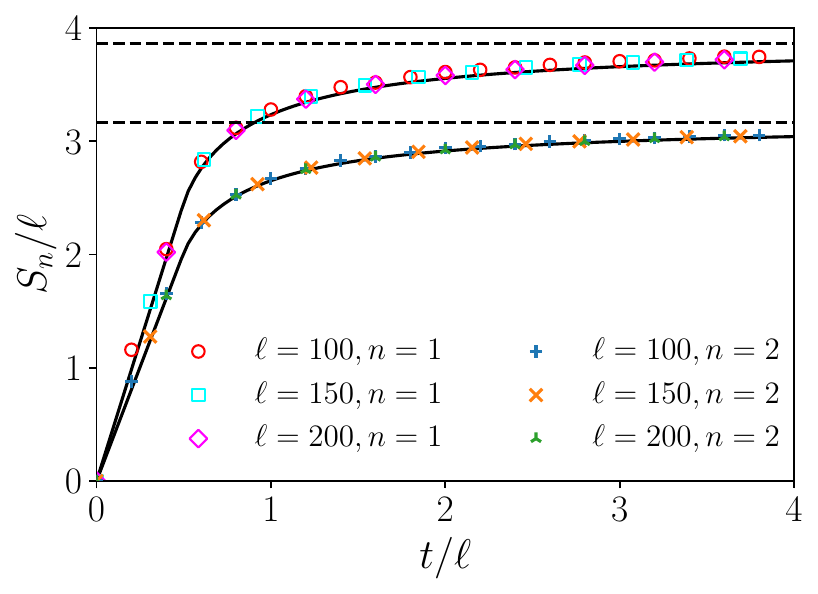}
    \caption{Time evolution of the R\'enyi entanglement entropy taking as initial configuration the diagonal dimer state. We consider different R\'enyi indices $n$ and subsystem sizes $\ell$. The solid lines represent the analytic expression in Eq.\,\eqref{eq:S_n_asymptotic_DD}. 
    The symbols are the exact numerical entropy computed with Eq.~\eqref{eq:S_n=tr_ln_Gamma} The dashed lines correspond to the saturation value predicted in Eq.\,\eqref{eq:Sn_saturation_DD}. 
    We set $L_y=10$.}
    \label{fig:Sn_DD}
\end{figure}

Substituting it into Eq.\,\eqref{eq:S_n_k-site} with $k=1$, we obtain that the entanglement entropy behaves as 
\begin{equation}
    S_n 
    \simeq 
    \sum_{q_y}
    \int_0^{2\pi}
    \frac{\dd q_x}{2\pi}
    h_n(\cos(q_x+q_y))
    \min(\ell,2t|v_x(q_x)|). 
    \label{eq:S_n_asymptotic_DD}
\end{equation}
Note that, unlike the previous cases, the contribution to the entropy of each mode $q_y$ is different. At large times, $t\gg \ell/2$, the entropy saturates to
\begin{align}
    \lim_{t\rightarrow\infty}
    S_n
    &\simeq 
    \ell 
    \sum_{q_y}
    \int_0^{2\pi}
    \frac{\dd q_x}{2\pi}
    h_n(\cos(q_x+q_y)). 
    \label{eq:Sn_saturation_DD}
\end{align}

In Fig.~\ref{fig:Sn_DD}, we analyze the entanglement entropy in the quantum quench from the diagonal dimer state. We obtain an excellent agreement between the analytic result of Eq.~\eqref{eq:S_n_asymptotic_DD} and the numerical values computed using Eq.~\eqref{eq:S_n=tr_ln_Gamma}.

\subsection{Crossed dimer state}
\label{subsec:CD}

Here we consider the quantum quench starting from the crossed dimer state
\begin{multline}
    \ket{\mathrm{C}}
    =
    \prod_{i_x=0}^{\frac{L_x}{2}-1}
    \prod_{i_y=0}^{\frac{L_y}{2}-1}
    \frac{1}{2}
    (a_{2i_x,2i_y}^\dag-a_{2i_x+1,2i_y+1}^\dag)
    \\
    \times
    (a_{2i_x+1,2i_y}^\dag-a_{2i_x,2i_y+1}^\dag)
    \ket{0},
    \label{eq:|C>}
\end{multline}
which is schematically illustrated in Fig.\,\ref{fig:initial_states} (f). Since this configuration is invariant under two-site translations in the $y$-direction, the entanglement entropy after the quench can be calculated by evaluating the moments $\Tr\,[(\Gamma_{q_y}^{(2)})^{2m}]$. To adapt the computation to the two-site translation symmetry in the $x$-direction, we rearrange the entries of $\Gamma_{q_y}^{(2)}$ as in Eq.\,\eqref{eq:Gamma_qy_rearrange}. This matrix can
be calculated employing Eqs.~\eqref{eq:cdaggerc_corr} and \eqref{eq:cc_corr} with the initial state 
two-point spatial correlators $\langle a_{\bf j}^\dagger a_{\bf j'}\rangle$ and $\langle a_{\bf j} a_{\bf j'}\rangle$. For the crossed dimer state, 
the latter read
\begin{widetext}
\begin{align}
    \bra{\rm C} a_{\bf j}^\dag a_{\bf j'}
    \ket{\rm C}
    =&
    \frac{\delta_{\bf j,j'}}{2}
    -
    \frac{1}{8}
    (\delta_{j_x+1,j_x'}\delta_{j_y+1,j_y'}
    +\delta_{j_x-1,j_x'}\delta_{j_y-1,j_y'}
    +\delta_{j_x-1,j_x'}\delta_{j_y+1,j_y'}
    +\delta_{j_x+1,j_x'}\delta_{j_y-1,j_y'})
    \nonumber \\
    &-
    \frac{(-1)^{j_x}}{8}
    (\delta_{j_x+1,j_x'}\delta_{j_y+1,j_y'}
    -\delta_{j_x-1,j_x'}\delta_{j_y-1,j_y'}
    -\delta_{j_x-1,j_x'}\delta_{j_y+1,j_y'}
    +\delta_{j_x+1,j_x'}\delta_{j_y-1,j_y'})
    \nonumber \\
    &-
    \frac{(-1)^{j_y}}{8}
    (\delta_{j_x+1,j_x'}\delta_{j_y+1,j_y'}
    -\delta_{j_x-1,j_x'}\delta_{j_y-1,j_y'}
    +\delta_{j_x-1,j_x'}\delta_{j_y+1,j_y'}
    -\delta_{j_x+1,j_x'}\delta_{j_y-1,j_y'})
    \nonumber \\
    &-
    \frac{(-1)^{j_x+j_y}}{8}
    (\delta_{j_x+1,j_x'}\delta_{j_y+1,j_y'}
    +\delta_{j_x-1,j_x'}\delta_{j_y-1,j_y'}
    -\delta_{j_x-1,j_x'}\delta_{j_y+1,j_y'}
    -\delta_{j_x+1,j_x'}\delta_{j_y-1,j_y'}),
\end{align}
\end{widetext}
and $\bra{{\rm C}} a_{\bf i} a_{{\bf i}'}\ket{{\rm C}}=0$.
We then obtain that $\Gamma_{q_y}^{(2)}$ is of the form
\begin{multline}\label{eq:Gamma_qy_CD}
    (\Gamma_{q_y}^{(2)})_{i_x,i_x'}
    =
    U
    e^{\im t \sigma_z\otimes \sigma_z \otimes I\cos q_y}
    \\
    \mqty(- \hat m(q_y) \otimes T_{i_x',i_x}^{\rm C}&0\\
    0&\hat m(q_y) \otimes T_{i_x,i_x'}^{\rm C})
    \\
    e^{-\im t \sigma_z\otimes \sigma_z\otimes I \cos q_y}
    U^\dag,
\end{multline}
where $U=I_8/2-\sum_{\mu=x,y,z}\sigma_\mu\otimes I_2\otimes \sigma_\mu/2$. The matrix $\hat m_{q_y}$ is defined as  
\begin{align}
    \hat m_{q_y}
    = 
    -\sigma_z \cos q_y
    +\sigma_y 
    \sin q_y,
\end{align}
and, in the thermodynamic limit $L_x\to\infty$, $T^{\rm C}$ reads
\begin{align}
    T_{i_x,i_x'}^{\rm C}
    = 
    \int_0^{2\pi}
    \frac{\dd q_x}{2\pi}
    e^{-2\im (i_x-i_x')}
    g_{\rm C}(q_x). 
\end{align}
Here the symbol $g_{\rm C}(q_x)$ is given by 
\begin{align}
    g_{\rm C}(q_x)
    = 
    e^{\im \frac{q_x}{2}\sigma_z}
    (
    \sigma_x \cos q_x
    +
    \sigma_+\sin q_x 
    e^{-2\im t \cos q_x}
    )
    e^{-\im \frac{q_x}{2}\sigma_z}.
\end{align}
\par 
Since the trace of the tensor product of two matrices is the product of the traces of the matrices, the moments $\Tr\,[(\Gamma_{q_y}^{(2)})^{2m}]$ in the present case are given by 
\begin{align}
    \Tr[(\Gamma_{q_y}^{(2)})^{2m}]
    = 
    2\Tr [\hat m_{q_y}^{2m}]
    \Tr [(T^{\rm C})^{2m}].
\end{align}
By simple algebra, one finds $\Tr[\hat m_{q_y}^{2m}]=2$ while, by employing the stationary phase method of Ref.\,\cite{Calabrese-2012}, one can also obtain the asymptotic form of $\Tr[(T^{\rm C})^{2m}]$ in the ballistic limit,
\begin{multline}
    \Tr [(T^{\rm C})^{2m}]
    \simeq 
    \ell 
    -
    \int_0^{2\pi}
    \frac{\dd q_x}{2\pi}
    [1-(\cos q_x)^{2m}]
    \\
    \times
    \min(\ell,2t|v_x(q_x)|). 
\end{multline}
Therefore, putting the previous results together, the moments of the matrix~\eqref{eq:Gamma_qy_CD} are
\begin{multline}
    \Tr[(\Gamma_{q_y}^{(2)})^{2m}]
    \simeq 
    4\ell
    -4 
    \int_0^{2\pi}
    \frac{\dd q_x}{2\pi}
    [1-(\cos q_x)^{2m}]
    \\
    \times
    \min(\ell,2t|v_x(q_x)|)
\end{multline}
and, plugging them in Eq.\,\eqref{eq:S_n_k-site} with $k=2$, we finally find that 
\begin{gather}
    S_n 
    = 
    L_y \int_0^{2\pi} \frac{\dd q_x}{2\pi}
    h_n(\cos q_x) \min(\ell,2t|v_x(q_x)|). 
    \label{eq:Sn_CD}
\end{gather}
In particular, the stationary value of the entanglement entropy at large times, $t\gg \ell/2$ is 
\begin{gather}
    \lim_{t\rightarrow \infty}S_n
    \simeq 
    V_A 
    \int_0^{2\pi}
    \frac{\dd q_x}{2\pi}
    h_n(\cos q_x). 
    \label{eq:S_n(inf)_CD}
\end{gather}

\begin{figure}
    \includegraphics[width=0.45\textwidth]{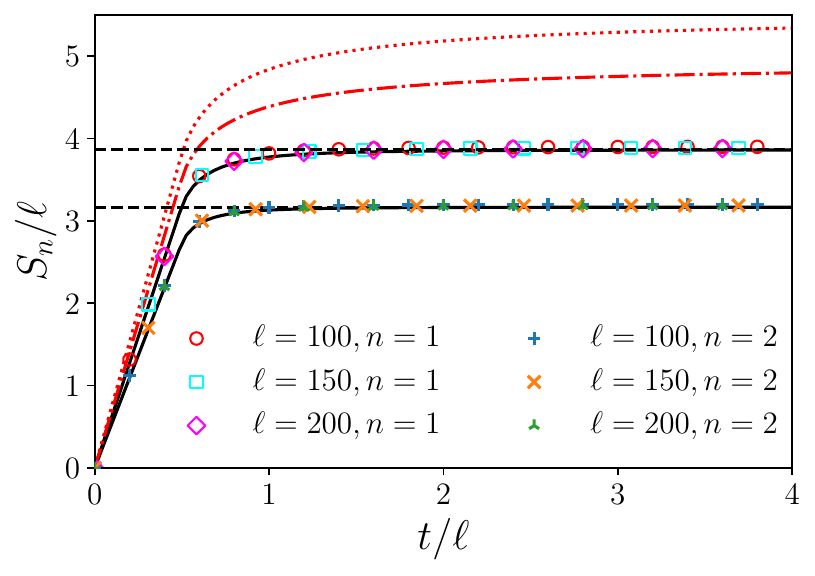}
    \caption{R\'enyi entanglement entropy as a function of $t/\ell$ in a quench from the crossed dimer state for R\'enyi indices $n\to 1$ and $n=2$ and different subsystem sizes $\ell$.
    The solid lines represent the analytic expression obtained in 
    Eq.\,\eqref{eq:Sn_CD} while the symbols are the exact value of the entropies calculated numerically using Eq.~\eqref{eq:S_n=tr_ln_Gamma}. The dashed lines are the saturation values at large times given by Eq.~\eqref{eq:S_n(inf)_CD}. 
    The dotted and dash-dotted lines are the predictions of the quasiparticle picture in Eq.\,\eqref{eq:Sn_CD_QP} for $n\to 1$ and $2$, respectively (see the discussion in the next section for the explanation of the disagreement). 
    In all the cases, we take $L_y=10$.} 
    \label{fig:Sn_CD}
\end{figure}

In Fig.~\ref{fig:Sn_CD}, we plot the entanglement entropy in the quantum quench starting from the crossed dimer state.  
It shows that the analytic expression obtained in Eq.\,\eqref{eq:Sn_CD} agrees well with the exact result obtained numerically with Eq.~\eqref{eq:S_n=tr_ln_Gamma}.

\subsection{Partially-filled product state}

So far, we have considered initial states with a defined number of particles, i.e. eigenstates of the particle number operator $Q=\sum_{{\rm i}}a_\mathbf{i}^\dag a_\mathbf{i}$. Let us now study quenches from configurations that break this $U(1)$ symmetry. 

We can construct them using as building block the 1D product state
\begin{equation}\label{eq:pf_bb}
\ket{\theta}_{i_y}=\prod_{i_x=0}^{L_x-1}
    \qty(
    \sin \frac{\theta}{2}
    +
    \cos \frac{\theta}{2}
    a_\mathbf{i}^\dag 
    )\ket{0}_{i_y},
\end{equation}
where $\ket{0}_{i_y}=\bigotimes_{i_x=0}^{L_x-1}\ket{0}_\mathbf{i}$ with $\ket{0}_\mathbf{i}$ being the local vacuum state for the $\mathbf{i}$-th site (in 1D spin language this is a tilted ferromagnetic state). 
The angle $\theta\in[0,\pi)$ controls the probability of finding a particle at the site ${\bf i}$ and, therefore, tunes how much the particle number symmetry is broken~\cite{Ares-2023}. At $\theta=0$ ($\pi$), the state~\eqref{eq:pf_bb} is fully-occupied (empty) and it preserves this $U(1)$ symmetry, whereas it breaks it for $\theta\neq 0,\pi$; in particular, the symmetry is maximally broken at $\theta=\pi/2$. 

The state~\eqref{eq:pf_bb} does not satisfy Wick theorem and, therefore, its reduced density matrix is not Gaussian and we cannot
calculate the entanglement entropy from the two-point correlation matrix using Eq.~\eqref{eq:S_n=tr_ln_Gamma}. However, the \textit{cat}
version of~\eqref{eq:pf_bb},
\begin{align}\label{eq:1dTF}
    \ket{\mathrm{PF}}_{i_y}
    =
    \frac{1}{\sqrt{2+2(\cos\theta)^{L_x}}}
    \left(\ket{\theta}_{i_y}-
    \ket{-\theta}_{i_y}\right).  
\end{align}
does satisfies Wick theorem and its reduced density matrix is Gaussian, as explicitly shown, e.g., in Ref.~\cite{amvc-23}.

From the 1D state $\ket{{\rm PF}}_{i_y}$, we can construct two 
different initial configurations in the two-dimensional square lattice, 
for which the entanglement entropy evolves differently after a quench
to $H$. We study them separately in the next subsections. 

\subsubsection{Partially-filled product state I}\label{subsec:TTF}
Let us first consider the state
\begin{align}
    \ket{\mathrm{PF}_{\rm I}}
    =\bigotimes_{i_y=0}^{L_y-1}
    \ket{\mathrm{PF}}_{i_y}, 
    \label{eq:|psi(0)>_TF}
\end{align}
where $\ket{\mathrm{PF}}_{i_y}$ is defined in Eq.~\eqref{eq:pf_bb}.
Since~\eqref{eq:|psi(0)>_TF} is invariant under one-site translations in the $y$-direction, we can employ Eq.\,\eqref{eq:S_n_k-site} with $k=1$ to calculate the evolution of the entanglement entropy after the quench. The entries of the correlation matrix $\Gamma_{q_y}^{(1)}$ that enters in such equation can be determined
using Eqs.~\eqref{eq:cdaggerc_corr} and \eqref{eq:cc_corr} with the initial values of the
two-point spatial correlators. In this case, given the product state
structure of the initial configuration~\eqref{eq:|psi(0)>_TF}, we have
\begin{equation}
\bra{{\rm PF}_{\rm I}}a_{\bf i}^\dagger a_{{\bf i}'}\ket{{\rm PF}_{\rm I}}={}_{i_y}\!\!\bra{{\rm PF}} a_{i_x, i_y}^\dagger a_{i_x', i_y}\ket{{\rm PF}}_{i_y}\delta_{i_y, i_y'}\,
\end{equation}
and 
\begin{equation}
\bra{{\rm PF}_{\rm I}}a_{\bf i} a_{{\bf i}'}\ket{{\rm PF}_{\rm I}}={}_{i_y}\!\!\bra{{\rm PF}} a_{i_x, i_y} a_{i_x', i_y}\ket{{\rm PF}}_{i_y}\delta_{i_y, i_y'}.
\end{equation}
The correlators for the 1D state $\ket{{\rm PF}}_{i_y}$ have been 
calculated in, e.g. Ref.~\cite{Ares-2023}. Using them here, we have
\begin{multline}
\bra{{\rm PF}_{\rm I}}a_{\bf i}^\dagger a_{{\bf i}'}\ket{{\rm PF}_{\rm I}}=\frac{\delta_{i_x, i_x'}\delta_{i_y, i_y'}}{2}
\\
-\frac{\delta_{i_y, i_y'}}{2L_x}\sum_{q_x} 
e^{-{\rm i}q_x(i_x-i_x')} \cos \Delta_{q_x}
\end{multline}
and
\begin{equation}
\bra{{\rm PF}_{\rm I}}a_{\bf i} a_{{\bf i}'}\ket{{\rm PF}_{\rm I}}=
-\frac{{\rm i}\delta_{i_y, i_y'}}{2L_x}\sum_{q_x}e^{-{\rm i}q_x(i_x-i_x')}\sin \Delta_{q_x},
\end{equation}
with
\begin{align}
    \cos \Delta_{q_x}
    &= 
    \frac{2|\cos \theta|-\cos q_x(1+\cos^2 \theta)}
    {1+\cos^2\theta-2|\cos \theta|\cos q_x},
    \label{eq:cos_Delta}
    \\
    \sin \Delta_{q_x}
    &=
    \frac{-\sin^2\theta\sin q_x}
    {1+\cos^2\theta-2|\cos \theta|\cos q_x}.
    \label{eq:sin_Delta}
\end{align}
Inserting them in Eqs.~\eqref{eq:cdaggerc_corr} and \eqref{eq:cc_corr}
and taking the limit $L_x\to\infty$, we find that $\Gamma_{q_y}^{(1)}$
is a block Toeplitz matrix,
\begin{align} (\Gamma_{q_y}^{(1)})_{i_x,i_x'}
    = 
    \int_0^{2\pi}
    \frac{\dd q_x}{2\pi}
    e^{-\im q_x(i_x-i_x')}
    g_{q_y}^{\rm PF}(q_x), 
\end{align}
generated by the $2\times 2$ symbol $g_{q_y}^{\rm PF}(q_x)$
\begin{align}
    g_{q_y}^{\rm PF}(q_x)
    = 
    \sigma_z \cos \Delta_{q_x}+ 
    \sigma_y e^{-2\im t \epsilon_\mathbf{q}\sigma_z}\sin \Delta_{q_x}.
    \label{eq:g_qy(q_x)_TTF}
\end{align}

The moments $\Tr\,[(\Gamma_{q_y}^{(1)})^{2m}]$ of a block Toeplitz matrix can be calculated by applying directly the stationary phase method of Ref.~\cite{Calabrese-2012}. In our case, we get
\begin{multline}
    \Tr[\Gamma_{q_y}^{2m}]
    \simeq 
    2\ell 
    -2\int_0^{2\pi}
    \frac{\dd q_x}{2\pi}
    [1-(\cos \Delta_{q_x})^{2m}]
    \\
    \times
    \min(\ell,2t|v_x(q_x)|).
\end{multline}
Plugging it into Eq.\,\eqref{eq:S_n_k-site} with $k=1$, we finally obtain that  
\begin{align}
    S_n 
    \simeq 
    L_y 
    \int_0^{2\pi}
    \frac{\dd q_x}{2\pi}
    h_n(\cos \Delta_{q_x})
    \min(\ell,2t|v_x(q_x)|).
    \label{eq:Sn_TTF_SP_limit}
\end{align}
In this case, the entropy converges at large times $t\gg\ell/2$ to the value
\begin{align}
    \lim_{t\rightarrow\infty}
    S_n 
    \simeq 
    V_A 
    \int_0^{2\pi}
    \frac{\dd q_x}{2\pi}
    h_n(\cos \Delta_{q_x}).
    \label{eq:Sn_TTF_saturation}
\end{align}

\begin{figure}
    \centering
    \includegraphics[width=0.5\textwidth]{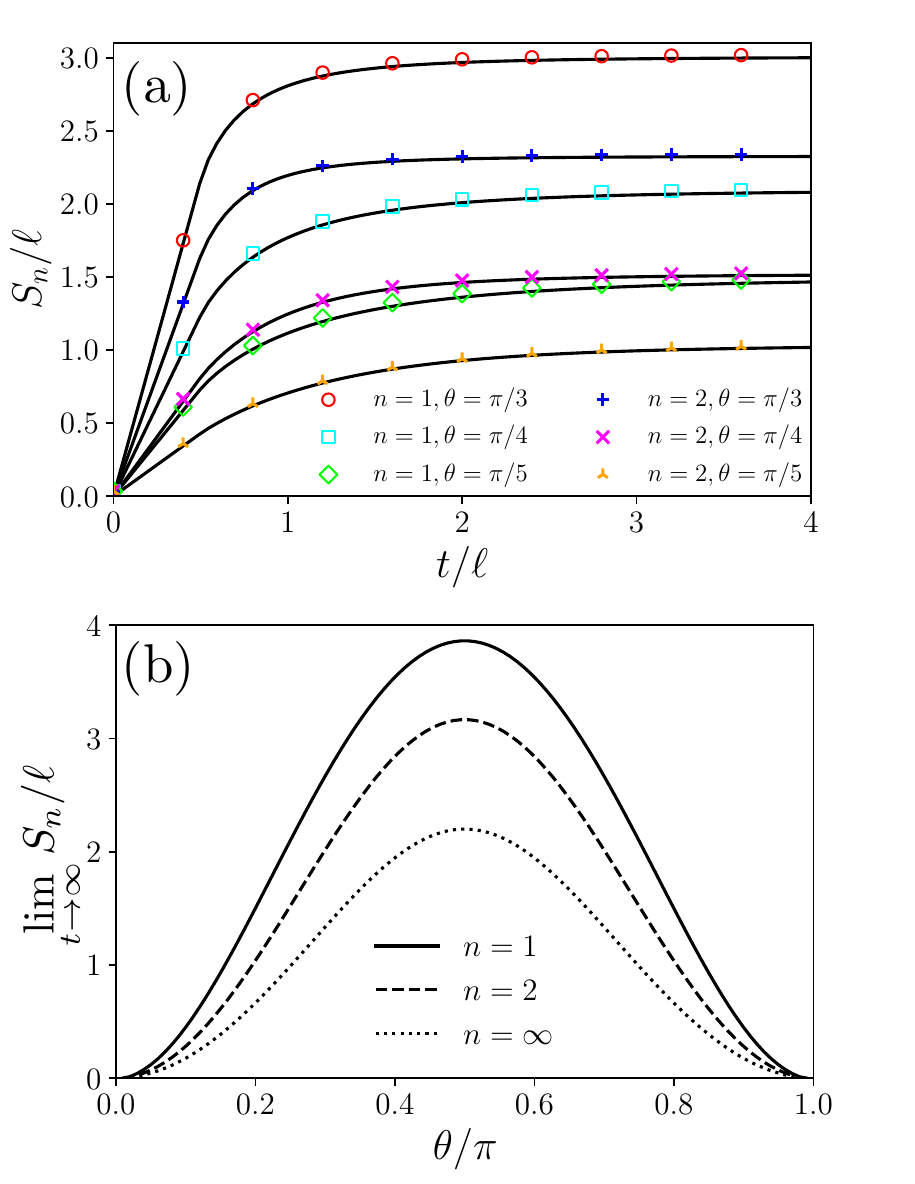}
    \caption{(a) Time evolution of the R\'enyi entanglement entropy after a quench from the state $\ket{{\rm PF}_{\rm I}}$, considering different angles $\theta$ for the initial configuration and R\'enyi indices $n$. The solid lines correspond to the analytic prediction found in Eq.\,\eqref{eq:Sn_TTF_SP_limit}. The symbols are the exact values computed numerically with Eq.~\eqref{eq:S_n=tr_ln_Gamma}.
    (b) Saturation value at large times of the R\'enyi entanglement entropy given in Eq.\,\eqref{eq:Sn_TTF_saturation} as a function of the initial angle $\theta$ for different R\'enyi indices $n$. In all cases,  we set $\ell=200$ and $L_y=10$.}
    \label{fig:TF_Sn}
\end{figure}

In Fig.\,\ref{fig:TF_Sn} (a), we plot the time evolution of the entanglement entropy after the quench from the state $\ket{{\rm  PF}_{\rm I}}$ for different values of the angle $\theta$ and the Rényi index $n$. We find that Eq.\,\eqref{eq:Sn_TTF_SP_limit} agrees well with the numerical value of the entropy calculated with Eq.~\eqref{eq:S_n=tr_ln_Gamma}. In Fig.~\ref{fig:TF_Sn} (b), we represent the stationary value of the entropy given by Eq.~\eqref{eq:Sn_TTF_saturation} as a function of the 
initial angle $\theta$. It can be seen that the entanglement entropy monotonically increases as $\theta$ increases until $\theta=\pi/2$, the angle at which the initial state \eqref{eq:|psi(0)>_TF} maximally breaks the $U(1)$ particle number symmetry.
This maximum value for the von Neumann entropy is 
$S_1^{\rm max}= V_A(2
\log (2)-1)$.

\subsubsection{Partially-filled product state II}

\begin{figure}
    \centering
    \includegraphics[width=0.5\textwidth]{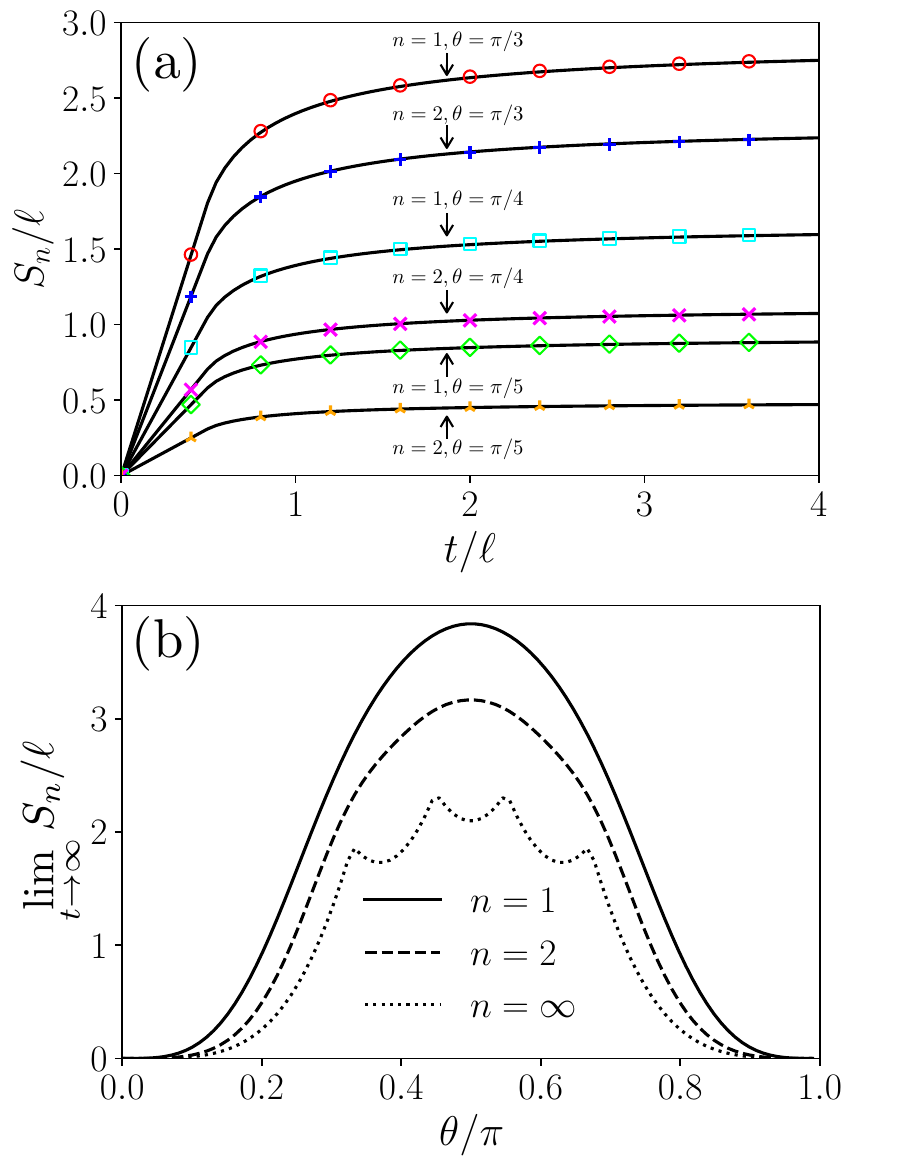}
    \caption{(a) R\'enyi entanglement entropy as a function of $t/\ell$ after a quench from the state $\ket{{\rm PF}_{\rm II}}$. We take several angles $\theta$ in the initial state and R\'enyi indices $n\to 1$ and $n=2$. The solid lines are the analytic expression derived in Eq.\,\eqref{eq:Sn_LTF} while the symbols represent the exact result obtained numerically employing Eq.~\eqref{eq:S_n=tr_ln_Gamma}
    (b) We represent the saturation value at large times of R\'enyi entanglement entropy predicted in Eq.\,\eqref{eq:Sn_LTF_saturation} versus the initial angle $\theta$. In all cases, we fix $\ell=200$ and $L_y=10$.}
    \label{fig:LTF_Sn}
\end{figure}

From the 1D state~\eqref{eq:pf_bb}, we can also construct the 
configuration
\begin{align}
    \ket{\mathrm{PF}_{\rm II}}
    = 
    \bigotimes_{i_x=0}^{L_x-1}
    \ket{\mathrm{PF}}_{i_x}.
\end{align}
The difference compared to state \eqref{eq:|psi(0)>_TF} is that this is a product state in the $x$-direction while the other was in the $y$-direction. 
Since the native 1D state \eqref{eq:1dTF} is a cat and not a product state, the two definitions are inequivalent.
This state $ \ket{\mathrm{PF}_{\rm II}}$ is invariant under one site translations in the $y$ and $x$ directions.  Since it is a product state along the $x$ direction
its two-point spatial correlation functions satisfy
\begin{equation}
\bra{{\rm PF}_{\rm II}}a_{\bf i}^\dagger a_{{\bf i}'}\ket{{\rm PF}_{\rm II}}={}_{i_x}\!\!\bra{{\rm PF}} a_{i_x, i_y'}^\dagger a_{i_x, i_y'}\ket{{\rm PF}}_{i_x}\delta_{i_x, i_x'}\,
\end{equation}
and 
\begin{equation}
\bra{{\rm PF}_{\rm II}}a_{\bf i} a_{{\bf i}'}\ket{{\rm PF}_{\rm II}}={}_{i_x}\!\!\bra{{\rm PF}} a_{i_x, i_y} a_{i_x, i_y'}\ket{{\rm PF}}_{i_x}\delta_{i_y, i_x'}.
\end{equation}
As we did in the other partially-filled product state, we can apply
the results of Ref.~\cite{Ares-2023} for the correlations of the 1D state $\ket{{\rm PF}}_{i_x}$. Then we have
\begin{multline}
\bra{{\rm PF}_{\rm II}}a_{\bf i}^\dagger a_{{\bf i}'}\ket{{\rm PF}_{\rm II}}=\frac{\delta_{i_x, i_x'}\delta_{i_y, i_y'}}{2}
\\
-\frac{\delta_{i_x, i_x'}}{2L_y}\sum_{q_y} 
e^{-{\rm i}q_y(i_y-i_y')} \cos \Delta_{q_y}
\end{multline}
and
\begin{equation}
\bra{{\rm PF}_{\rm II}}a_{\bf i} a_{{\bf i}'}\ket{{\rm PF}_{\rm II}}=
-\frac{{\rm i}\delta_{i_x, i_x'}}{2L_y}\sum_{q_y}e^{-{\rm i}q_y(i_y-i_y')}\sin \Delta_{q_y}.
\end{equation}

Inserting these correlators in Eqs.~\eqref{eq:cdaggerc_corr} and \eqref{eq:cc_corr} and taking the thermodynamic limit $L_x\rightarrow\infty$, the matrix $\Gamma_{q_y}^{(1)}$ that gives the entanglement entropy~\eqref{eq:S_n_k-site} is block Toeplitz, 
\begin{gather}
    (\Gamma_{q_y}^{(1)})_{i_x,i_x'}
    = 
    \int_0^{2\pi}
    \frac{\dd q_x}{2\pi}
    e^{-\im q_x(i_x-i_x')}
    g_{q_x}^{\rm PF}(q_y), 
    \label{eq:Gamma_qy_LTF}
\end{gather}
in which the symbol is the same as in Eq.~\eqref{eq:g_qy(q_x)_TTF} but exchanging
the moments $q_x$ and $q_y$.
\par 
As in the previous examples, in the ballistic limit, the asymptotic form of the moments $\Tr\,[(\Gamma_{q_y}^{(1)})^{2m}]$ for Eq.\,\eqref{eq:Gamma_qy_LTF} can be obtained by applying the stationary phase method for block Toeplitz matrices of Ref.\,\cite{Calabrese-2012}. Here we find
\begin{multline}
    \Tr[(\Gamma_{q_y}^{(1)})^{2m}]
    \simeq 
    2\ell 
    -2[1-(\cos \Delta_{q_y})^{2m}]
    \\
    \times
    \int_0^{2\pi}
    \frac{\dd q_x}{2\pi}
    \min(\ell,2t|v_x(q_x)|).
\end{multline}
Applying this result in Eq.~\eqref{eq:S_n_k-site} with $k=1$, we get that the entanglement entropy evolves in time after the quench as
\begin{gather}
    S_n 
    \simeq 
    \sum_{q_y}
    h_n(\cos \Delta_{q_y})
    \int_0^{2\pi}
    \frac{\dd q_x}{2\pi}
    \min(\ell,2t|v_x(q_x)|),
    \label{eq:Sn_LTF}
\end{gather}
and saturates to 
\begin{align}
    \lim_{t\rightarrow\infty}
    S_n
    =
    \ell
    \sum_{q_y}
    h_n(\cos \Delta_{q_y}) 
    \label{eq:Sn_LTF_saturation}
\end{align}
at large times $t\gg \ell/2$.
\par 
Fig.~\ref{fig:LTF_Sn} (a) shows the entanglement entropy of the time-evolved state $\ket{\mathrm{PF}_{\rm II}(t)}$, comparing the analytic result of Eq.~\eqref{eq:Sn_LTF} (solid lines) with  the exact value (symbols) obtained numerically with Eq.~\eqref{eq:S_n=tr_ln_Gamma} for several initial angles $\theta$ and Rényi index $n$. 
In Fig.\,\ref{fig:LTF_Sn} (b), we plot the saturation value of the entanglement entropy found in Eq.\,\eqref{eq:Sn_LTF_saturation} as a function of $\theta$. We can see that, as in the case of the state $\ket{{\rm PF}_{\rm I}}$, the saturation value is maximum at $\theta=\pi/2$ (at least for $n=1,2$), at which the particle number symmetry is maximally broken at $t=0$, although it is not in general monotonic in $\theta$.
The nonanalytic behavior for $n=\infty$ in Fig.\,\ref{fig:LTF_Sn} (b) disappears when $L_y$ is sufficiently large.

\section{Quasiparticle picture}
\label{sec:quasiparticle_picture}

In this section, we discuss the physical interpretation of the results
of the previous section in terms of the {\it quasiparticle picture}, originally developed for one-dimensional integrable systems~\cite{Calabrese-2005, Alba-2017}. 

The idea of the quasiparticle picture is the following: 
in one dimension the quench generates uniformly pairs of entangled quasiparticles that propagate ballistically in opposite directions with opposite momenta. 
Therefore, the entanglement at time $t$ after the quench is proportional to the number of pairs of entangled quasiparticles that are shared by subsystem $A$ and its complement $B$:
\begin{equation}
\label{eq:S_n_QP-1D}
S_n=\int_{0}^{2\pi} \frac{{\rm d}q}{2\pi} s_n( q) \min(\ell, 2|v(q)|t),
\end{equation} 
where $v(q)$ is the velocity of the quasiparticle with momentum $q$ and $s_n(q)$ its contribution to the entanglement. The function $\min(\ell, 2|v(q)|t)$ counts the quasiparticles that are shared by subsystem $A$ and its complement at time $t$.
\par
However, in $d>1$ as in the present paper, the multiplet structure of the quasiparticles is generically much more complicated than simple pairs and their counting is far from trivial. 
Fortunately, the dimensional reduction comes in our help, leading to extremely simple results.  
Indeed,  since the R\'enyi entanglement entropy can be decomposed in the single-interval entanglement entropies of decoupled 1D chains, we can directly apply the quasiparticle picture in each decoupled chain, which are labeled by the transverse momentum $q_y=2 \pi n_y/L_y$. 
Therefore, after the quench, in the $n_y$-chain pairs of entangled excitations propagate with momentum $\pm q_x$ and velocity $v_x(q_x)=\partial_{q_x}\epsilon_{{\bf q}}$. 
If we denote the contribution to the entanglement entropy of each pair of entangled modes as $s_n({\bf q})$ and sum over all the 
decoupled chains the result \eqref{eq:S_n_QP-1D}, then we expect
\begin{equation}\label{eq:S_n_QP-2}
S_n=\sum_{n_y=0}^{\frac{L_y}k-1}\int_{0}^{2\pi} \frac{{\rm d}q_x}{2\pi} s_n({\bf q}) \min(\ell, 2|v_x(q_x)|t).
\end{equation} 
Eq.~\eqref{eq:S_n_QP-2} predicts that the entanglement entropy increases linearly for $t< \ell/2$ and tends to a constant at large times $t\gg \ell/2$, this is the qualitative behavior that we have found in all the examples studied in Sec.\,\ref{sec:examples}.
 
To obtain quantitative predictions from Eq.\,\eqref{eq:S_n_QP-2}, we have to determine a specific expression of $s_n(\mathbf{q})$. 
For one-dimensional free-fermion systems, as proposed in Ref.~\,\cite{Alba-2017} for generic integrable systems, the analogous function $s_n({q})$ can be read off from the limit $t\rightarrow\infty$, in which the reduced density matrix $\rho_A(t)$ is expected to relax to a GGE. We can apply the same idea here. In particular, for the 2D free-fermion model \eqref{eq:H}, one may expect that the infinite time limit of $\rho_A(t)$ exists and it is described by 
a GGE, i.e.,
\begin{align}
    \lim_{t\rightarrow\infty}\rho_A(t)
    =
    \Tr_B(\rho^\mathrm{GGE})
    \equiv 
    \rho_{A}^\mathrm{GGE}.
    \label{eq:rho_A(inf)}
\end{align}
As in one dimension~\cite{fe-13}, since the post-quench Hamiltonian is diagonal in terms of the modes $\tilde a_{\bf q}^\dagger$ and $\tilde a_{\bf q}$, the GGE can be written in terms of the 
the conserved mode occupation numbers $\hat n_\mathbf{q}=\tilde a_\mathbf{q}^\dag \tilde a_\mathbf{q}$ as 
\begin{align}
    \rho^\mathrm{GGE}
    = 
    \frac{e^{-\sum_\mathbf{q}\lambda_\mathbf{q} \hat n_\mathbf{q}}}
    {\Tr(e^{-\sum_\mathbf{q}\lambda_\mathbf{q}\hat n_\mathbf{q}})},
    \label{eq:rho_GGE}
\end{align}
where the Lagrange multipliers $\lambda_{\bf q}$ are determined  by the expectation value of $\hat n_\mathbf{q}$, 
\begin{align}
    \Tr(\hat n_\mathbf{q} \rho^\mathrm{GGE})
    =
    \bra{\psi_0}
    \hat n_\mathbf{q}
    \ket{\psi_0}
    \equiv n_\mathbf{q}. 
    \label{eq:tr(n_rho_GGE)}
\end{align}
We stress that this is valid under the quite general assumption that non-Abelian charges like $\sum_j (-1)^j c_j c_{j+m}$ are not activated by the initial state; their activation would lead to an altered dynamics \cite{fagotti,amvc-23}. 
\par 
If the reduced density matrix $\rho_A(t)$ relaxes to the GGE as in Eq.\,\eqref{eq:rho_A(inf)}, then the Rényi entanglement entropy for $\rho_A(t)$ at large times must be equal to the entropy of the statistical ensemble $\rho_{A}^\mathrm{GGE}$,
\begin{align}
    \lim_{t\rightarrow\infty}
    S_n(\rho_A(t))
    = 
    S_n(\rho_A^{\mathrm{GGE}}).
    \label{eq:S_n=S_GGE_n}
\end{align}
\par 
We can deduce the specific form of $s_n(\mathbf{q})$ from the above equation as follows.
For a large subsystem with volume $V_A$, $S_{n}(\rho^{\mathrm{GGE}}_{A})$ is proportional to $V_A$ because it is an extensive thermodynamic quantity. 
Hence the entropy $S_{n}(\rho^{\mathrm{GGE}}_{A})$ is given by the volume of subsystem $A$ times the density of the Rényi entropy evaluated in the GGE of Eq.~\eqref{eq:rho_GGE}. That is, 
\begin{align}
    S_n(\rho_{A}^\mathrm{GGE})
    &= 
    \frac{V_A}{L_xL_y}
    S_n(\rho^\mathrm{GGE})\nonumber
    \\
    &= 
    \ell 
    \sum_{n_y=0}^{L_y-1}
    \int_0^{2\pi}
    \frac{\dd q_x}{2\pi}
    h_n(2n_\mathbf{q}-1),
    \label{eq:Sn(rho_GGE,A)}
\end{align}
where the function $h_n(x)$ is defined in Eq.~\eqref{eq:h_n}. 
In the second line, we have taken the thermodynamic limit $L_x\rightarrow\infty$ to derive it. 
On the other hand, if we take the large time limit $t\rightarrow\infty$ in Eq.\,\eqref{eq:S_n_QP-2}, we obtain 
\begin{align}
    \lim_{t\rightarrow\infty} S_n(\rho_A)
    =
    \ell \sum_{n_y=0}^{L_y/k-1}
    \int_0^{2\pi}
    \frac{\dd q_x}{2\pi}
    s_n(\mathbf{q}). 
    \label{eq:Sn(rho_A)_lim_QP}
\end{align}
Comparing Eqs.~\eqref{eq:Sn(rho_GGE,A)} and \eqref{eq:Sn(rho_A)_lim_QP} and naturally grouping together the chains with equal $n_y$ modulo $k$, we can conclude that 
\begin{equation}
s_n(\mathbf{q})=\sum_{j=0}^{k-1}h_n\left(2n_{q_x, q_y+\frac{2\pi j}{k}}-1\right).
\label{eq:s_n=h_n}
\end{equation}
Finally, plugging this result into Eq.\,\eqref{eq:S_n_QP-2}, we find
\begin{align}
    S_n= 
    \sum_{n_y=0}^{L_y-1}
    \int_0^{2\pi}
    \frac{\dd q_x}{2\pi}
    h_n(2n_\mathbf{q}-1) 
    \min(\ell,2t|v_x(q_x)|). 
    \label{eq:RE_QP}
\end{align}
We emphasize that this prediction  assumes the thermodynamic limit in the $x$-direction, but the value of $L_y$ is an arbitrary integer, also as small as 1 or 2. 
\par 
The quasiparticle formula \eqref{eq:RE_QP} coincides with the analytic expressions found for most of the particular initial states investigated in Sec.~\ref{sec:examples}. 
For example, the mode occupation number for the collinear Mott insulator state \eqref{eq:|CAF>} is 
$n_\mathbf{q}=1/2$ and, substituting it in Eq.\,\eqref{eq:RE_QP}, we find Eq.\,\eqref{eq:S_n_2D_Neel_sp_limit}. 
Also occupation number for the Mott insulator state \eqref{eq:|NAF>} is $n_\mathbf{q}=1/2$. 
In a similar way, it is straightforward to check that the prediction of the quasiparticle picture \eqref{eq:RE_QP} reproduces the analytic expressions for the entanglement entropy obtained in Eqs.\,\eqref{eq:S1_sp_limit_vbs} (collinear and staggered dimer states), \eqref{eq:S_n_asymptotic_DD} (diagonal dimer state), \eqref{eq:Sn_TTF_SP_limit} (partially-filled product state I), and \eqref{eq:Sn_LTF} (partially-filled product state II). 
\par 
However, the situation is different for the crossed dimer state discussed in Sec.\,\ref{subsec:CD}. In this case, the mode 
occupation number is
\begin{align}
    n_\mathbf{q}
    =
    \bra{\mathrm{C}} \hat n_\mathbf{q}
    \ket{\mathrm{C}}
    = 
    \frac{1-\cos q_x\cos q_y}{2}
\end{align}
and, if we plug it in the quasiparticle prediction\,\eqref{eq:RE_QP}, then we obtain 
\begin{align}
    S_n 
    = 
    \sum_{n_y=0}^{L_y-1}
    \int_0^{2\pi}
    \frac{\dd q_x}{2\pi}
    h_n(\cos q_x\cos q_y)
    \min(\ell,2t|v_x(q_x)|),
    \label{eq:Sn_CD_QP}
\end{align}
which does not match the correct result in Eq.\,\eqref{eq:Sn_CD}, as also shown in Fig.\,\ref{fig:Sn_CD}. 
\par 
Comparing Eqs.\,\eqref{eq:Sn_CD} and \eqref{eq:Sn_CD_QP}, it is clear that the reason why the quasiparticle picture breaks down is that the identification \eqref{eq:s_n=h_n} does not hold. 
Since this identity has been derived assuming Eq.~\eqref{eq:S_n=S_GGE_n}, the mismatch means that, in this particular quench, the reduced density matrix $\rho_A(t)$ does not  tend to $\rho^{\mathrm{GGE}}_A$ at $t\to\infty$. In the following section, we will discern the reason why this occurs.

\section{Existence of stationary state} 
\label{sec:reduced_density_matrix}

In the previous section, we have found that the prediction of the quasiparticle picture \eqref{eq:RE_QP} does not match the correct result \eqref{eq:Sn_CD} for the crossed dimer state. 
We claimed that the reason for this disagreement is that the reduced density matrix $\rho_A(t)$ does not relax to the GGE $\rho_A^{\rm GGE}$ at large times. 
In this section, we show that, in two-dimensional free fermionic systems, the reduced density matrix $\rho_A(t)$ does not tend to a stationary 
state after quenches from certain particular initial configurations, including the crossed dimer state. 
Furthermore, we derive a criterion which allows us to know whether the stationary state exists or not for a given initial state and how the quasiparticle picture must be modified in its absence.

\subsection{Absence of stationary state}

To show that $\rho_A(t)$  does not relax to a stationary state for particular initial states, we start by decomposing the post-quench Hamiltonian \eqref{eq:H} as 
\begin{align}
    H 
    = 
    H^{X}
    + 
    H^{Y}, 
    \label{eq:H=H^x+H^y}
\end{align}
where 
\begin{align}
    H^{X}
    &=
    \sum_{i_y=0}^{L_y-1}
    H_{i_y}^{X},
    \nonumber\\
    &=
    -\frac{1}{2}
    \sum_{i_y=0}^{L_y-1}
    \sum_{i_x=0}^{L_x-1}
    a_{i_x+1,i_y}^\dag 
    a_{i_x,i_y}
    +\mathrm{H.c.},
    \label{def:H^X}
\end{align}
and
\begin{align}
    H^{Y}
    &=
    \sum_{i_x=0}^{L_x-1} H_{i_x}^{Y},
    \nonumber\\
    &=
    -\frac{1}{2}
    \sum_{i_x=0}^{L_x-1}
    \sum_{i_y=0}^{L_y-1}
    a_{i_x,i_y+1}^\dag 
    a_{i_x,i_y}
    +\mathrm{H.c.}. 
    \label{def:H^Y}
\end{align} 
As clear from the above equations, the terms $H^X$ and $H^Y$ only contain fermion hoppings in the $x$- and $y$-direction respectively. 
\par 
Observe that, in terms of the Fourier modes~\eqref{eq:b_modes}, $H^X$ and $H^Y$ are diagonal
\begin{equation}
H^X=-\sum_{{\bf q}}\cos q_x \tilde a_{{\bf q}}^\dagger \tilde a_{{\bf q}},
\end{equation}
\begin{equation}
H^Y=-\sum_{{\bf q}}\cos q_y \tilde a_{{\bf q}}^\dagger \tilde a_{{\bf q}},
\end{equation}
and, therefore, they commute with each other. This allows us to write the time-evolved state $\ket{\psi(t)}$ as 
\begin{align}
    \ket{\psi(t)}
    = 
    e^{-\im t H^Y}
    \ket{\psi_X(t)}, 
\end{align}
where $\ket{\psi_X(t)}=e^{-\im tH^X}\ket{\psi_0}$ is the time-evolved state after a quantum quench with a Hamiltonian in which the fermion hopping is allowed only in the $x$-direction (i.e., $L_y$ copies of 1D chains).      
Accordingly, the reduced density matrix $\rho_A(t)$ can be written as 
\begin{align}
    \rho_A(t) 
    = 
    \Tr_B\left( e^{-\im t H^Y} \ket{ \psi_X(t)}
    \bra{\psi_X(t)} e^{\im t H^Y}\right). 
    \label{eq:rho_A(t)}
\end{align}
Given its definition in Eq.~\eqref{def:H^Y}, we have $[H_{i_x}^Y,H_{i_x'}^Y]=0$; then the time-evolution operator $e^{-\im t H^Y}$ in Eq.~\eqref{eq:rho_A(t)} can be decomposed into two operators that respectively act only on the subsystems $A$ and $B$ as 
\begin{align}
    e^{-\im t H^Y}
    = 
    U_A(t)
    U_B(t),
    \label{eq:e^itH^y_decompose}
\end{align}
with 
\begin{align}
    U_A(t)=\prod_{i_x=0}^{\ell-1}e^{-\im t H^Y_{i_x}},\ 
    U_B(t)=\prod_{i_x=\ell}^{L_x-1}e^{-\im t H^Y_{i_x}}.
    \label{eq:U_A}
\end{align}
Note that $U_A$ acts only on subsystem $A$ and hence it can be taken out the partial trace $\Tr_B$. 
Therefore, we can rewrite Eq.\,\eqref{eq:rho_A(t)} as 
\begin{align}
    \rho_A(t) 
    = 
    U_A(t) \rho_{X,A}(t)U_A^\dag(t),  
    \label{eq:rho_A=Urho_XAU}
\end{align}
where 
\begin{align}
    \rho_{X,A}(t) 
    = 
    \Tr_B\ket{\psi_X(t)}\bra{\psi_X(t)}.
    \label{eq:rho_XA}
\end{align}
Note that, in the latter expression, $U_B$ and $U_B^\dag$ cancel each other in the partial trace $\Tr_B$ due to the cyclic property.
From Eq.~\eqref{eq:rho_A=Urho_XAU}, we conclude that, when $[U_A(t),\rho_{X,A}(\infty)]\neq 0$, the limit $\lim_{t\rightarrow\infty}\rho_A(t)$ does not exist, i.e., there is no stationary state. 

\subsection{Condition for the existence of stationary state}
\label{subsec:condition_stationary_state}

Having established that the existence of a stationary state is related to the 
vanishing of the commutator $[U_A, \rho_{X, A}(\infty)]$, 
the next natural step is to determine under which conditions this commutator  is not zero. 
In Appendix~\ref{app:proof_of_theorem}, 
we prove the following. If we assume that $\rho_{X, A}(t)$ relaxes to a 
stationary state, i.e. $\lim_{t\to\infty}\rho_{X, A}(t)$ exists, then 
the stationary state of $\rho_A(t)$ exists if and only if $\rho_{X, A}(t)$
restores the translational symmetry in the $y$-direction for large
times, that is
\begin{align}
    \mathcal{T}_A \qty(\lim_{t\rightarrow\infty}\rho_{X,A}(t)) \mathcal{T}_A^{-1}
    =
    \lim_{t\rightarrow\infty}\rho_{X,A}(t), 
    \label{eq:Trho_XT}
\end{align}
where $\mathcal{T}_A$ is the translation operator in the $y$-direction
action on subsystem $A$.

\begin{figure}
    \centering
    \includegraphics[width=0.5\textwidth]{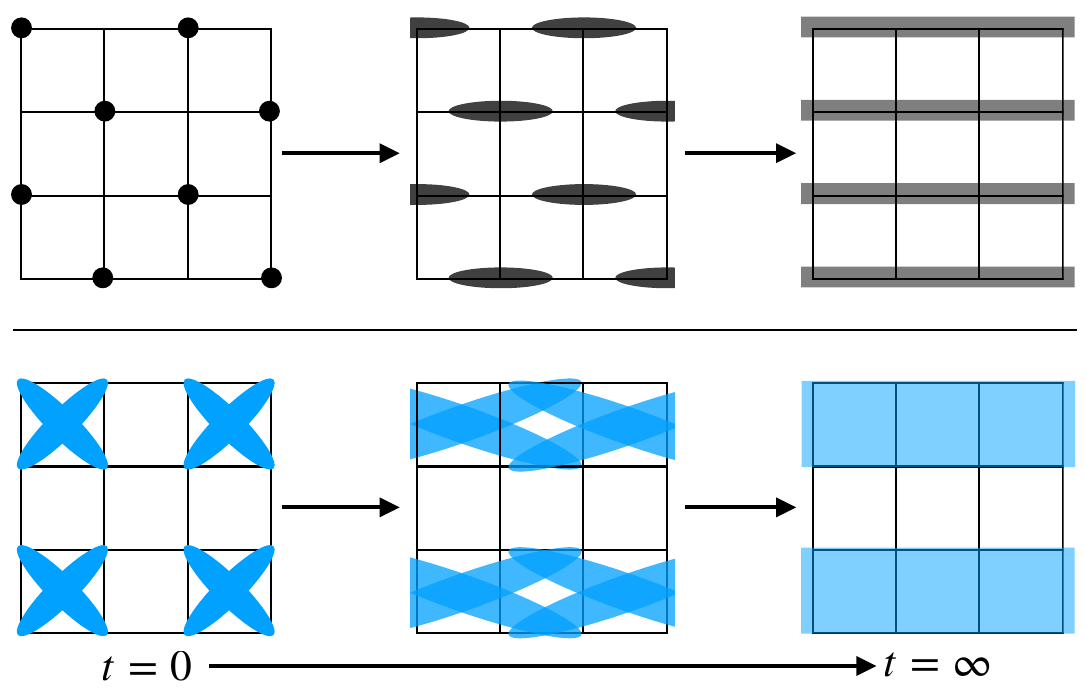}
    \caption{Schematic representation of time evolution of the reduced density matrix $\rho_{X,A}(t)$, introduced in Eq.~\eqref{eq:rho_XA}, after quenches from the Mott insulator state (upper panel) and the crossed dimer state (lower panel).}
    \label{fig:rho_XA}
\end{figure}

This result allows us to know whether the stationary state exists or 
not from $\rho_{X,A}$. For example, in Fig.~\ref{fig:rho_XA}, we schematically represent the time evolution of $\rho_{X,A}(t)$ in quenches starting from the Mott insulator and the crossed dimer states. 
In the case of the Mott insulator state, after a long time, the hopping in the $x$-direction makes the distribution of fermions uniform for every $i_y$-th row, and the translational symmetry in the $y$-direction is restored. Hence the stationary state exists according to the 
result announced before. On the other hand, for the crossed dimer state, since the hopping in the $x$-direction does not change the amount of entanglement between each row, after long times, the entanglement between $2i_y$-th  and $(2i_y+1)$-th rows in the initial state remains, while no entanglement appears between $(2i_y-1)$-th and $2i_y$-th rows. This means that $\rho_{X,A}$ for the crossed dimer state never restores the translational symmetry in the $y$-direction. 
Therefore, the reduced density matrix $\rho_{A}(t)$ does not relax to a stationary state in this case.  

In general, as we show in Appendix~\ref{app:lemma}, the density 
matrix $\rho_{X, A}(t)$ restores the translational invariance in the 
$y$-direction at large times, i.e. Eq.~\eqref{eq:Trho_XT} is 
satisfied, if and only if 
\begin{align}
        \bra{\psi_0} \tilde a_{q_x,q_y}^\dag \tilde a_{q_x,q_y'}\ket{\psi_0}
        =0\ \forall q_x,q_y\neq q_y'.
        \label{eq:lemma}
    \end{align} 
Therefore, we can conclude that $\rho_A(t)$ relaxes to a stationary state at large times if and only if the initial state satisfies Eq.~\eqref{eq:lemma}.
In fact, one can check that such correlator is always zero for all the initial states discussed in Sec.~\ref{sec:examples} except for the crossed dimer state, for which we have
\begin{align}
    \bra{\rm C} \tilde a_{q_x,q_y}^\dag \tilde a_{q_x,q_y+\pi} \ket{\rm C}
    =
    -\frac{\im }{2}\cos q_x \sin q_y.
\end{align}

\subsection{Entanglement entropy in the absence of stationary state: the crossed dimer state} \label{subsec:crossed-dimer_2}

Although, as we have just seen, there is not a stationary state in a 
quench from the crossed dimer state, in Sec.~\ref{sec:examples} we showed that the entanglement entropy saturates to a constant value at large times. 
This behavior can be explained as follows.
According to Eq.~\eqref{eq:rho_A=Urho_XAU}, the reduced density matrices $\rho_A(t)$ and $\rho_{X, A}(t)$ are related by a unitary transformation. This means
that their entropies are equal,
\begin{align}
     S_n(\rho_A(t)) 
     = 
     S_n(\rho_{X,A}(t)). 
     \label{eq:Sn(rho_A)=Sn(rho_XA)}
 \end{align}
Therefore, this identity implies that, if $\rho_{XA}(t)$ relaxes to a stationary state, then the entropy tends to a constant value at large times, even if the limit $\lim_{t\to\infty}\rho_A(t)$ does not exist. 

The missing point is to determine the stationary value of the entropy in the absence of stationary state for $\rho_A(t)$. We have shown that, if $[U_A, \rho_{X, A}(\infty)]=0$, then both $\rho_{A}(t)$ and $\rho_{X, A}(t)$ relax to the same stationary state, which is described by a GGE that can be built either from the local conserved charges of $H=H^X+H^Y$ or of $H^X$. However, the set of conserved charges of $H^X$ is not equal to that of the total
Hamiltonian $H$. Therefore, if $\rho_A(t)$ tends to a stationary state, only the charges shared by $H^X$ and $H$ can be activated. On the contrary, if there is not stationary state for $\rho_A(t)$, it exists the possibility that charges only conserved by $H^X$ can be activated. In such cases, the entanglement entropies derived from the GGEs associated to $H$ and $H^X$ are different, and
the correct stationary value of the entanglement entropy is given by the latter. This explains why the quasiparticle picture discussed in Sec.~\ref{sec:quasiparticle_picture} does not work for the crossed dimer state, for which $\rho_A(t)$ does not relax as we have already shown.

In fact, for this initial state, we find that, to obtain the saturation value of Eq.~\eqref{eq:S_n(inf)_CD} at large times, the correct GGE should be built with the following conserved charges
\begin{align}
     \hat n_{q_x,i_y}^\pm 
     \equiv 
     \frac{1}{\sqrt{2}}
     (d_{q_x,2i_y}^\dag\pm d_{q_x,2i_y+1}^\dag )
     (d_{q_x,2i_y}\pm d_{q_x,2i_y+1}), 
 \end{align}
where $q_x=0,\dots, 2\pi(L_x-1)/L_x$, $i_y=0,\dots, L_y/2-1$, and $\hat d_{q_x,i_y}$ is the partial Fourier transform of the fermionic $a_\mathbf{i}$ with respect to the $x$-direction  
\begin{align}
     d_{q_x,i_y}
     \equiv  
     \frac{1}{\sqrt{L_x}}
     \sum_{i_x=0}^{L_x-1}
     e^{-\im q_x i_x }
     a_\mathbf{i}.
 \end{align}
By simple algebra, one finds that $\{\hat n_{q_x,i_y}^\pm\}$ commute with $H^X$ while they do not commute with the total Hamiltonian $H=H^X+H^Y$. 

 In the same way as in Eq.~\eqref{eq:Sn(rho_GGE,A)}, we can calculate the entropy of a GGE built with the charges $\{n_{q_x, y_i}^\pm\}$. In the thermodynamic limit $L_x\to \infty$, it is
 \begin{equation}
   S_n
    = 
     \ell 
     \sum_{i_y=0}^{\frac{L_y}{2}-1}
     \int_0^{2\pi}
     \frac{\dd q_x}{2\pi}
     [
     h_n(2n_{q_x,i_y}^+-1)
     +
     h_n(2n_{q_x,i_y}^--1)
     ],
     \label{eq:Sn(rho_GGEXA)}
 \end{equation}
where we have introduced $n_{q_x,i_y}^\pm=\bra{\psi_0}\hat n_{q_x,i_y}^\pm \ket{\psi_0}$. In particular, for the crossed dimer state, we obtain 
 \begin{align}
     n_{q_x,i_y}^\pm
     =
     \bra{\mathrm{C}}
     \hat n_{q_x,i_y}^\pm
     \ket{\mathrm{C}}
     =
     \frac{1\pm\cos q_x}{2}.
 \end{align}
Plugging it into Eq.\,\eqref{eq:Sn(rho_GGEXA)}, we indeed recover the large time limit of the evolution from the crossed  dimer state in Eq.\,\eqref{eq:S_n(inf)_CD}.
At this point, Eq. \eqref{eq:Sn_CD} for the finite time evolution is recovered by a straightforward application of the quasiparticle picture. 

In Appendix~\ref{subsec:ITF}, we consider a quench from another initial configuration in which $\rho_A(t)$ does not relax to a stationary state and show that, also in this case, the large time behavior of the entanglement entropy is captured by a GGE built with from the charges of $H^X$ instead of $H$.

\section{Conclusions}
\label{sec:summary}

To summarize, we studied the time evolution of the entanglement entropy following quantum quenches in a translationally invariant 2D free-fermion lattice. 
We have considered different initial Gaussian configurations that present a periodic pattern in both directions of the system. 
By applying dimensional reduction and exploiting the 
well-known results for 1D free fermionic chains, we analytically determined the exact behavior of the R\'enyi entanglement entropies after the quench. 
We found that, for most of the initial states, the standard quasiparticle picture developed for 1D systems can be readily adapted to correctly explain the evolution of the entropy. 
Instead, for one particular initial state, the quasiparticle picture does not generalize straightforwardly because the R\'enyi entanglement entropies saturate to
a value different from the one 
predicted by the GGE built with the local conserved charges of the post-quench Hamiltonian. 
We traced back the origin of this disagreement to the absence of a stationary limit for the reduced density matrix. 
Starting from this observation, we deduced that the correct stationary entanglement entropy for this initial state is given by the GGE constructed with the local conserved charges of a Hamiltonian with only hoppings in the longitudinal direction. We also obtained the general conditions for the existence of the stationary limit of the reduced density matrix, which are related to the restoration of the translational symmetry in the transverse direction at large times.

The dimensional reduction approach employed here can be easily generalized to study the 
time evolution in two-dimensional free fermionic systems of other quantities 
for which there are 
exact results in one dimension. These include
the entanglement negativity~\cite{ac-19, ctc-14, mac-21,pbc-22}, charge fluctuations \cite{gec-17,bcckr-22}, symmetry-resolved entanglement entropies~\cite{parez2021, parez2021b,pvcc-22, bcckr-22, bkccr-23}, and the 
recently introduced entanglement asymmetry~\cite{Ares-2023, amvc-23, ckacmb-23, fac-23}.
The latter measures how much a symmetry is broken
in a subsystem, and has been employed to observe a quantum version~\cite{Ares-2023}
of the Mpemba effect \cite{cool,kumar, luraz, raz2}. A relevant question is whether this effect may also occur in higher dimensions and under what conditions it happens.
Another easy but interesting generalization would be to use dimensional reduction to study quench problems in 2D free bosonic models to contrast with existing field theory literature \cite{cotler2016}.

\acknowledgements
We thank V.~Alba, S.~Murciano, and C.~Rylands for fruitful discussions. 
SY is supported by Grant-in-Aid for JSPS Fellows (Grant No.\,JP22J22306).
PC and FA acknowledge support from ERC under Consolidator Grant number
771536 (NEMO).
We thank the authors of Ref. \cite{gjsb-23} for sharing their results with us before submission.

\section*{Note Added}

After the completion of this work we became aware of the parallel work \cite{gjsb-23}, where also the entanglement entropy of 2D free fermion systems is studied. However, in this manuscript the dimensional reduction is not used and the emphasis is more on the shape on the entangling region rather than on the different initial states. 

\appendix

\section{Proof of the condition of Eq.~\eqref{eq:Trho_XT} for the existence of stationary state}
\label{app:proof_of_theorem}

In this Appendix, we prove that reduced density matrix $\rho_A(t)$ relaxes to a stationary state if and only if Eq.\,\eqref{eq:Trho_XT} holds. 
In other words, the goal of the appendix is to prove that the following two propositions are necessary and sufficient conditions for each other: 
\begin{prop}
\label{p1}
The limit  
\[\lim_{t\rightarrow\infty} \rho_A(t)\]
exists. 
\end{prop}
\begin{prop}
\label{p2}
For large times, $\rho_{X,A}(t)$ in Eq. \eqref{eq:rho_XA} restores the translational symmetry in the $y$-direction, i.e., the following equation holds:  
\[\mathcal{T}_A \rho_{X,A}(\infty) \mathcal{T}_A^{-1}=\rho_{X,A}(\infty).\] 
\end{prop}
To this end, we make the following assumptions: 
\begin{asm}
\label{a1}
$\rho_{X,A}(t)$ relaxes to the GGE associated to $H^X$ (cf. Eq. \eqref{def:H^X}) at large time, i.e., 
\begin{align}
    \lim_{t\rightarrow\infty} \rho_{X,A}(t)
    = 
    \rho_{X,A}^{\rm GGE}.
\end{align}
\end{asm}
\begin{asm}
\label{a2}
If the stationary state for $\rho_A(t)$ exists, then 
\begin{align}
    \lim_{t\rightarrow\infty} \rho_{A}(t)
    = 
    \rho_{A}^{\rm GGE},
\end{align}
where $\rho_A^{\rm GGE}$ is the GGE built from the conserved charges of $H$.
\end{asm}
Note that we can use Asm.\,\ref{a2} only when Prop.\,\ref{p1} is true. 

\par 
To prove Prop.\,\ref{p1} $\Longleftrightarrow$ Prop.\,\ref{p2}, we first show that Prop.\,\ref{p1} is true if and only if the following proposition is true:
\begin{prop}
\label{p3}
$\rho_{X,A}(t)$ converges to $\rho_{A}^\mathrm{GGE}$ for large times, i.e., 
\[\rho_{X,A}^\mathrm{GGE}=\rho_A^\mathrm{GGE}.\]
\end{prop}
\par
\underline{Proposition \ref{p3} $\Longrightarrow$ Prop.\,\ref{p1}} can be shown as follows: from Eq.\,\eqref{eq:rho_A=Urho_XAU}, 
\begin{align}
    \lim_{t\rightarrow\infty}
    \rho_{A}(t)
    =
    \lim_{t\rightarrow\infty}
    U_A(t) \rho_{A,X}(t)U_A^\dag(t).  
\end{align}
Assumption \ref{a1} and Prop.\,\ref{p3} allow us to replace $\rho_{A,X}$ in the above equation with $\rho_{A}^\mathrm{GGE}$, which reads  
\begin{align}
    \lim_{t\rightarrow\infty}
    \rho_{A}(t)
    =
    \lim_{t\rightarrow\infty}
    U_A(t) \rho_{A}^\mathrm{GGE}U_A(t)^\dag
    =
    \rho_A^\mathrm{GGE}. 
\end{align}
Here we have used $[U_A(t),\rho_A^\mathrm{GGE}]=0$ in the last equality. 
The above equation clearly shows that
\begin{align}
    \mathrm{Prop.\,\ref{p3}}\Longrightarrow\mathrm{Prop.\,\ref{p1}}.
    \label{eq:p3_to_p1}
\end{align}
\par 
\underline{Proposition \ref{p1} $\Longrightarrow$ Prop.\,\ref{p3}} can also be shown in a similar way: from Eq.\,\eqref{eq:rho_A=Urho_XAU}, 
\begin{align}
    \lim_{t\rightarrow\infty} \rho_{X,A}(t)
    =
    \lim_{t\rightarrow\infty} U_A^\dag(t) \rho_A(t) U_A(t). 
\end{align}
Proposition \ref{p1} and Asm.\,\ref{a2} allow us to replace $\rho_A$ in the above equation with $\rho_A^\mathrm{GGE}$, which results in 
\begin{align}
    \lim_{t\rightarrow\infty} \rho_{X,A}(t)
    =
    \lim_{t\rightarrow\infty} U_A^\dag(t) \rho_A^\mathrm{GGE} U_A(t)
    =
    \rho_A^\mathrm{GGE}. 
\end{align}
Here we again used $[U_A(t),\rho_A^\mathrm{GGE}]=0$ in the last equality. From the previous equation, we can conclude that
\begin{align}
    \mathrm{Prop.\,\ref{p1}}\Longrightarrow\mathrm{Prop.\,\ref{p3}}.
    \label{eq:p1_to_p3}
\end{align}
\par 
Eqs. \eqref{eq:p3_to_p1} and \eqref{eq:p1_to_p3} imply
\begin{align}
    \mathrm{Prop.}\,\ref{p1}\Longleftrightarrow\mathrm{Prop.}\,\ref{p3}. 
    \label{(p1)_iff_(p3)}
\end{align}
\par 
By \eqref{(p1)_iff_(p3)}, proving Prop.\,\ref{p1} $\Longleftrightarrow$ Prop.\,\ref{p2} is equivalent to show Prop.\,\ref{p2} $\Longleftrightarrow$ Prop.\,\ref{p3}. Let us therefore prove the latter. 
\par 
\underline{Proposition\, \ref{p3} $\Longrightarrow$ Prop.\,\ref{p2}} can be shown as follows: from Prop.\,\ref{p3}, we obtain 
\begin{align}
    \mathcal{T}_A \rho_{X,A}^\mathrm{GGE}\mathcal{T}_A^{-1}
    &=
    \mathcal{T}_A \rho_{A}^\mathrm{GGE}\mathcal{T}_A^{-1},
    \\
    &=
    \Tr_B[\mathcal{T}_A \rho^\mathrm{GGE}\mathcal{T}_A^{-1}]. 
\end{align}
Inserting the identity $\mathcal{T}_B^{-1} \mathcal{T}_B$ in $\Tr_B$ in the above equation and using the cyclic property of trace, we obtain 
\begin{align}
    \mathcal{T}_A \rho_{X,A}^\mathrm{GGE}\mathcal{T}_A^{-1}
    =
    \Tr_B[\mathcal{T}_B \mathcal{T}_A 
    \rho^\mathrm{GGE}
    \mathcal{T}_A^{-1}
    \mathcal{T}_B^{-1} ].
    \label{eq:TrT}
\end{align}
According to Eq.\,\eqref{eq:rho_GGE}, $\rho^\mathrm{GGE}$ is invariant under translations in the $y$-direction, namely, 
\begin{align}
    \mathcal{T}_B \mathcal{T}_A 
    \rho^\mathrm{GGE}
    \mathcal{T}_A^{-1}
    \mathcal{T}_B^{-1}
    =
    \rho^\mathrm{GGE}. 
\end{align}
Substituting it into Eq.\,\eqref{eq:TrT}, we obtain 
\begin{align}
    \mathcal{T}_A \rho_{X,A}^\mathrm{GGE}\mathcal{T}_A^{-1}
    =
    \rho^\mathrm{GGE}_A
\end{align}
and using Prop.\,\ref{p3}
\begin{align}
    \mathcal{T}_A \rho_{X,A}^\mathrm{GGE}\mathcal{T}_A^{-1}
    =
    \rho^\mathrm{GGE}_{X,A}. 
\end{align}
The above equation shows that 
\begin{align}
    \mathrm{Prop.\,}\ref{p3}\Longrightarrow\mathrm{Prop.\,}\ref{p2}.
    \label{eq:p3_to_p2}
\end{align}
\par 
\underline{Proposition\,\ref{p2} $\Longrightarrow$ Prop.\,\ref{p3}} can be shown as follows: 
Without loss of generality, $\rho_{X,A}^\mathrm{GGE}$ can be expressed as 
\begin{align}\label{eq:ent_ham_XA}
    \rho_{X,A}^\mathrm{GGE}
    =
    \frac{e^{-\mathcal{H}_A}}{\Tr_A(e^{-\mathcal{H}_A})}, 
\end{align}
where $\mathcal{H}_A$ is the entanglement Hamiltonian which is given by~\cite{peschel-02} 
\begin{align}
    \mathcal{H}_A
    =
    \sum_{\mathbf{i,i'}\in A} K_{\mathbf{i,i'}} a_{\bf i}^\dag a_{\bf i'}, 
\end{align}
with $K$ being a $V_A\times V_A$ Hermitian matrix. 
Therefore, Prop.\,\ref{p2} is equivalent to stating that the entanglement Hamiltonian $\mathcal{H}_A$ is invariant under translations in the $y$-direction, i.e., 
\begin{align}
    \mathrm{Prop.\,\ref{p2}}\Longleftrightarrow \mathcal{T}_A \mathcal{H}_A\mathcal{T}_A^{-1}
    =\mathcal{H}_A. 
\end{align}
This implies that we can decompose $\mathcal{H}_A$ in the transverse momentum sectors by taking the partial Fourier transform~\eqref{def:c_ix_qy} in the $y$-direction,
\begin{align}
    \mathcal{H}_A
    =
    \sum_{q_y} \sum_{i_x,i_x'=0}^{\ell-1}
    [K(q_y)]_{i_x,i_x'} c_{i_x,q_y}^\dag c_{i_x',q_y}, 
    \label{eq:H_bd}
\end{align}
where $K(q_y)$ is a $\ell\times\ell$ Hermitian matrix. 
On the other hand, if we express the matrix $U_A(t)$, defined in 
Eq.~\eqref{eq:U_A}, in the mixed space-momentum basis,
\begin{align}
    U_A(t) =e^{-\im t \sum_{i_x=0}^{\ell-1} \sum_{q_y}\cos q_y c_{i_x,q_y}^\dag  c_{i_x,q_y}},
\end{align}
we find that
\begin{align}
    U_A(t) c_{i_x,q_y} U_A^\dag(t) 
    = 
    c_{i_x,q_y} e^{-\im t \cos q_y}. 
\end{align}
Combining this result with Eq.\,\eqref{eq:H_bd}, we obtain 
\begin{align}
    U_A(t) \mathcal{H}_A U_A^\dag(t) =\mathcal{H}_A. 
\end{align}
Therefore, applying this identity in Eq.~\eqref{eq:ent_ham_XA}, 
\begin{align}
    U_A(t) \rho_{A,X}^\mathrm{GGE}U_A^\dag(t) = \rho_{A,X}^\mathrm{GGE}. 
    \label{eq:UrU=r}
\end{align}
Thus, if we take the limit
\begin{align}
    \lim_{t\rightarrow\infty} 
    \rho_A(t) 
    =
    \lim_{t\rightarrow\infty} 
    U_A(t) \rho_{X,A}(t) U_A^\dag(t),
    \label{eq:rho_A=rho_AX^GGE_}
\end{align}
we apply Asm.~\ref{a1},
\begin{align}
    \lim_{t\to\infty}\rho_A(t)
    =
    \lim_{t\rightarrow\infty} 
    U_A(t) \rho_{X,A}^\mathrm{GGE} U_A^\dag(t),
\end{align}
and Eq.~\eqref{eq:UrU=r}, we obtain 
\begin{align}
    \lim_{t\to\infty}{\rho_A(t)}
    = 
    \rho_{X,A}^\mathrm{GGE}. 
    \label{eq:rho_A=rho_AX^GGE}
\end{align}  
Equation \eqref{eq:rho_A=rho_AX^GGE} shows that, if $\rho_{X, A}$ restores the transverse translational symmetry at large times, then $\rho_A(t)$ relaxes to a stationary state, i.e., Prop.\,\ref{p1} holds. Since we can use Asm.\,\ref{a2} when Prop.\,\ref{p1} holds, we can replace the left-hand side of Eq.\,\eqref{eq:rho_A=rho_AX^GGE} with $\rho_A^\mathrm{GGE}$ and
\begin{align}
    \rho_A^\mathrm{GGE} 
    =
    \rho_{X,A}^\mathrm{GGE}. 
\end{align}
Therefore,
$\mathrm{Prop.\,\ref{p2}}\Longrightarrow\mathrm{Prop.\,\ref{p3}}$.
This ends the proof of
\begin{align}
    \mathrm{Prop.\,\ref{p2}}\Longleftrightarrow\mathrm{Prop.\,\ref{p3}}. 
    \label{eq:p2_iff_p3}
\end{align} 
From \eqref{(p1)_iff_(p3)} and \eqref{eq:p2_iff_p3}, we have finally proved that 
\begin{align}
    \mathrm{Prop.\,\ref{p1}} \Longleftrightarrow \mathrm{Prop.\,\ref{p2}}.
    \label{p1_iff_p2_iff_p3}
\end{align}

\section{Proof of the condition~\eqref{eq:lemma} for the restoration
of translational invariance in the $y$-direction}
\label{app:lemma}
In this Appendix, we show that $\rho_{X,A}(t)$ restores the translation symmetry in the $y$-direction at large times, i.e., Prop.\,\ref{p2} is true, if and only if Eq.\,\eqref{eq:lemma} is satisfied. 
\par 
As mentioned in Appendix \ref{app:proof_of_theorem}, Prop.\,\ref{p2} is equivalent to saying that the stationary state $\lim_{t\rightarrow\infty}\rho_{A,X}(t)=\rho_{X,A}^\mathrm{GGE}$ is block diagonal in the  $q_y$ momentum sectors. 
Thus, we can verify whether Prop.\,\ref{p2} is satisfied or not by calculating the mixed space-momentum correlator
\begin{align}
    C_{i_x,i_x'}^{q_y,q_y'}(t)
    =
    \Tr_A(\rho_{X,A}(t) c_{i_x,q_y}^\dag c_{i_x',q_y'}).   
\end{align}
Given Asm.\,\ref{a1}, it is clear that $C_{i_x,i_x'}^{q_y,q_y'}(\infty)=0$ $\forall i_x,i_x'\in A, q_y\neq q_y'$ if and only if $\rho_{X,A}^\mathrm{GGE}$ is block diagonal in the $q_y$ momentum sectors. 
Thus, in the following, we prove that $C_{i_x,i_x'}^{q_y,q_y'}(\infty)=0$ if and only if Eq.\,\eqref{eq:lemma} holds. 
\par 
Using Asm.\,\ref{a1} and performing the Fourier transform in the $x$-direction, we can rewrite $C_{i_x,i_x'}^{q_y,q_y'}(\infty)$ as 
\begin{align}
    C_{i_x,i_x'}^{q_y,q_y'}(\infty)
    &=
    \frac{1}{L_x}\sum_{q_x,q_x'}
    e^{-\im q_x i_x+\im q_x'i_x'}
    \Tr(\rho^\mathrm{GGE}_X \Tilde{a}_\mathbf{q}^\dag \Tilde{a}_\mathbf{q'}).
    \label{C}
\end{align}
We recall that $\rho_X^\mathrm{GGE}$ is a GGE built with conserved charges of $H^X$, which is diagonalized as 
\begin{align}
    H^X
    &= 
    \sum_{q_x} \sum_{i_y} \cos q_x d_{q_x,i_y}^\dag d_{q_x,i_y},
    \\
    &= 
    \sum_{q_x} \sum_{\mu} \cos q_x \alpha_{q_x,\mu}^\dag \alpha_{q_x,\mu}, 
    \label{eq:H^X_diagonal}
\end{align}
where $\alpha_{q_x,\mu}=\sum_{i_y} [U(q_x)]_{\mu,i_y}d_{q_x,i_y}$ with $U(q_x)$ being a $L_y\times L_y$ unitary matrix. 
Therefore, without loss of generality, the conserved charges of $H^X$ are given by $\hat Q_{q_x,\mu}=\alpha_{q_x,\mu}^\dag \alpha_{q_x,\mu}$ and hence $\rho_X^\mathrm{GGE}$ can be written as 
\begin{align}
    \rho_X^\mathrm{GGE}
    &=
    \frac{e^{-\sum_{q_x,\mu}\lambda_{q_x,\mu}\hat Q_{q_x,\mu}}}
    {\Tr(e^{-\sum_{q_x,\mu}\lambda_{q_x,\mu}\hat Q_{q_x,\mu}})}.
\end{align}
Even without knowing explicitly the form of $U(q_x)$, the previous equation shows that $\rho_X^\mathrm{GGE}$ has a block-diagonal form in the $q_x$ momentum sectors. This implies that
\begin{align}
    \Tr(\rho^\mathrm{GGE}_X \tilde a_\mathbf{q}^\dag \tilde a_\mathbf{q'})
    &=
    \delta_{q_x,q_x'} 
    \Tr(\rho^\mathrm{GGE}_X \tilde a_{q_x,q_y}^\dag \tilde a_{q_x,q_y'}).
    \label{eq:tr_rho_X^GGE_bb}
\end{align}
Plugging it into Eq.\,\eqref{C}, we obtain 
\begin{align}
    C_{i_x,i_x'}^{q_y,q_y'}
    (\infty)
    =
    \frac{1}{L_x}
    \sum_{q_x} 
    e^{-\im q_x(i_x-i_x')}
    \Tr(\rho^\mathrm{GGE}_X \tilde a_{q_x,q_y}^\dag \tilde a_{q_x,q_y'}). 
    \label{eq:C_}
\end{align}
Since the charges $\tilde a_{q_x, q_y}^\dagger \tilde a_{q_x, q_y'}$ are conserved
by $H^X$, 
\begin{equation}
 \Tr(\rho^\mathrm{GGE}_X \tilde a_{q_x,q_y}^\dag \tilde a_{q_x,q_y'})=
  \bra{\psi_0} \tilde a_{q_x,q_y}^\dag \tilde a_{q_x,q_y'}\ket{\psi_0}
\end{equation} 
and, substituting it into Eq.\,\eqref{eq:C_}, we arrive at 
\begin{align}
    C_{i_x,i_x'}^{q_y,q_y'}
    (\infty)
    =
    \frac{1}{L_x}
    \sum_{q_x} 
    e^{-\im q_x(i_x-i_x')}
    \bra{\psi_0}
    \tilde a_{q_x,q_y}^\dag \tilde a_{q_x,q_y'}
    \ket{\psi_0}. 
\end{align}
From this expression, we find that $C_{i_x,i_x'}^{q_y,q_y'} (\infty)=0$ if and only if 
\begin{align}
    \bra{\psi_0}
     \tilde a_{q_x,q_y}^\dag \tilde a_{q_x,q_y'}
    \ket{\psi_0}=0\ \forall q_x, q_y\neq q_y'. 
\end{align}

\section{Inhomogeneous partially-filled product state}
\label{subsec:ITF}

In this Appendix, we analyze another example of a quench in which $\rho_A(t)$ does not relax to a stationary state, and thus the standard quasiparticle picture must be modified, see Sec. \ref{sec:reduced_density_matrix}. 
Specifically, we consider the quantum quench from the inhomogeneous partially-filled product state
\begin{align}
    \ket{\mathrm{IPF}} 
    = 
    \prod_{i_y=0}^{L_x-1}
    \ket{\rm IPF}_{i_y},
    \label{eq:|ITF>}
\end{align}
where $\ket{{\rm IPF}}_{i_y}$ is similar to the 1D cat state
of Eq.~\eqref{eq:1dTF} but now the angle $\theta$ that controls
the occupation probability depends on the coordinate $i_y$,
\begin{gather}
    \ket{\mathrm{IPF}}_{i_y}
    =
    \frac{1}{\sqrt{2+2(\cos\theta_{i_y})^{L_x}}}
   \left(\ket{\theta_{i_y}}-\ket{-\theta_{i_y}}\right).  
\end{gather} 
When $\theta_{i_y}=\theta$ for all $i_y$, $\ket{\mathrm{IPF}}$ is the the partially-filled product state I discussed in Sec.\,\ref{subsec:TTF}. 
\par 
Evaluating the correlator of Eq.~\eqref{eq:lemma} in the state $\ket{\rm IPF}$, we have
\begin{multline}
    \bra{\rm IPF} \tilde a_{q_x,q_y}^\dag \tilde a_{q_x,q_y'}\ket{\rm IPF}
    =\\=
    \frac{1}{2L_y}
    \sum_{i_y}
    e^{\im (q_y-q_y')i_y}
    (1-
    \cos \Delta_{q_x,i_y}),
    \label{eq:ITF_criterion}
    \end{multline}
where $\cos \Delta_{q_x,i_y}$ is obtained by replacing $\theta$ in Eq.\,\eqref{eq:cos_Delta} with $\theta_{i_y}$. 
If $\theta_{i_y}$ is independent of $i_y$, which corresponds to the case in Sec.\,\ref{subsec:TTF}, this correlator vanishes for all $q_x$ and $q_y\neq q_y'$. In that case, according to the results in Sec.\,\ref{subsec:condition_stationary_state},  $\rho_A(t)$ relaxes to a stationary state and the prediction of the standard quasiparticle picture \eqref{eq:RE_QP} works. 
On the other hand, when $\theta_{i_y}$ depends on $i_y$, the correlator of Eq.\,\eqref{eq:ITF_criterion} is in general nonzero, and Eq.\,\eqref{eq:lemma} is not satisfied. 
This implies that it does not exist a stationary state for $\rho_A(t)$.
\par  
When $\theta_{i_y}$ depends on $i_y$, the state $\ket{\mathrm{IPF}}$ has no translational symmetry in the $y$-direction, hence we cannot apply the dimensional reduction approach of Sec.~\ref{sec:dimension_reduction}. 
Instead, when the initial state is the product state for each $i_y$-th row as in Eq.\,\eqref{eq:|ITF>}, we can calculate the R\'enyi entanglement entropy without using the dimensional reduction as follows. Let us decompose the time evolution Hamiltonian as in Eq.~\eqref{eq:H=H^x+H^y}.
Since the operator $e^{-\im t H^X}$ has no dynamics in the $y$-direction, it preserves the initial product structure for each $i_y$-th row. Therefore, the reduced density matrix $\rho_{X,A}(t)$, introduced in Eq.~\eqref{eq:rho_XA}, and obtained in this case from the state $e^{-{\rm i}tH^X}\ket{\mathrm{IPF}}$, is of the form
\begin{align}
    \rho_{X,A}(t)
    = 
    \bigotimes_{i_y=0}^{L_y-1}
    \rho_{X,A,i_y}(t),
    \label{eq:rho_XA_ITF}
\end{align}
where
\begin{align}
    \rho_{X,A,i_y}(t)
    = 
    \Tr_{B,i_y}\left(
    e^{-\im t H_{i_y}^X} 
    \ket{\mathrm{IPF}}_{i_y}\!\!
    \bra{\mathrm{IPF}}
    e^{\im t H_{i_y}^X}\right), 
\end{align}
and $\Tr_{B,i_y}$ is the partial trace over the sites of the subsystem $B$ in the $i_y$-th row. Since $\rho_A(t)$ and $\rho_{X, A}(t)$ are related by the unitary transformation~\eqref{eq:rho_A=Urho_XAU}, we have $S_n(\rho_A(t))=S_n(\rho_{X, A}(t))$. Therefore, if we further take 
into account that the R\'enyi entanglement entropy is additive in the tensor product,
\begin{align}
    S_n(\rho_A) 
    = 
    S_n(\rho_{X,A})
    =
    \sum_{i_y=0}^{L_y-1}
    S_n(\rho_{X,A,i_y}). 
    \label{eq:RE_ITF}
\end{align}
\par 
Since $\rho_{X,A,i_y}$ is nothing but the reduced density matrix of the partially-filled product state I of Sec.~\ref{subsec:TTF} with $L_y=1$, the asymptotic form of $S_n(\rho_{X,A,i_y})$ in the space-time scaling limit can be obtained by just setting $L_y=1$ in Eq.\,\eqref{eq:Sn_TTF_SP_limit}. 
It reads 
\begin{align}
    S_n(\rho_{X,A,i_y})
    \simeq 
    \int_0^{2\pi}
    \frac{\dd q_x}{2\pi}
    h_n(\cos \Delta_{q_x,i_y})
    \min(\ell,2t|v_x(q_x)|).
    \label{eq:Sn_ITF_asymptotic}
\end{align}
Plugging the above equation into Eq.\,\eqref{eq:RE_ITF} yields 
\begin{align}
    S_n
    \simeq 
    \sum_{i_y=0}^{L_y-1}
    \int_0^{2\pi}
    \frac{\dd q_x}{2\pi}
    h_n(\cos \Delta_{q_x,i_y})
    \min(\ell,2t|v_x(q_x)|). 
    \label{eq:Sn_ITF_SP_limit}
\end{align}
\begin{figure}[t]
    \includegraphics[width=0.48\textwidth]{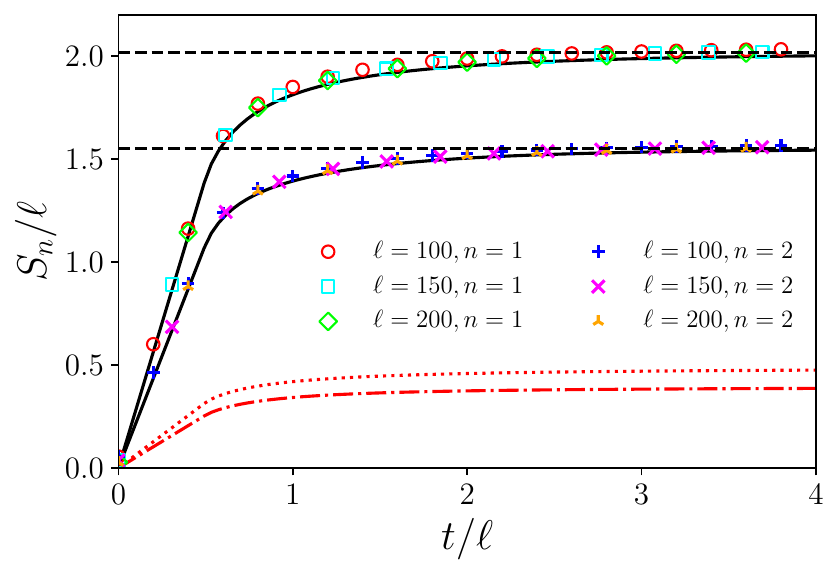}
    \caption{Time evolution of the R\'enyi entanglement entropy after a quench from the inhomogeneous partially-filled product state, taking as inhomogeneous angles $\theta_{i_y}=2\pi i_y/L_y$ for the initial configuration and R\'enyi indices $n\to 1$ and $n=2$.
    The solid lines correspond to the analytic prediction found in Eq.\,\eqref{eq:Sn_ITF_SP_limit}. 
    The symbols are the exact values of the entropy computed numerically with Eq.\,\eqref{eq:S_n=tr_ln_Gamma}. 
    The dashed lines are the large-time saturation value predicted in Eq.\,\eqref{eq:Sn_ITF_SP_limit}. 
    The dotted and dash-dotted lines are the predictions of the standard quasiparticle picture in Eq.\,\eqref{eq:Sn_ITF_QP} for $n\rightarrow1$ and $n=2$, respectively. 
    In all cases, we take $L_y=10$.}
    \label{fig:Sn_ITF}
\end{figure}
Accordingly, the saturation value of the R\'enyi entanglement entropy is
\begin{align}
    \lim_{t\rightarrow\infty}
    S_n
    &=
    \ell 
    \sum_{i_y=0}^{L_y-1}
    \int_0^{2\pi}
    \frac{\dd q_x}{2\pi}
    h_n(\cos \Delta_{q_x,i_y}). 
    \label{eq:Sn_ITF_t_limit}
\end{align}
On the other hand, applying the standard quasiparticle picture of Sec.~\ref{sec:quasiparticle_picture} to the state $\ket{\mathrm{IPF}}$, i.e. Eq.~\eqref{eq:RE_QP} with 
the expectation value 
of the conserved charges~\eqref{eq:ITF_criterion}, we find 
\begin{multline}
    S_n
    \simeq 
    L_y
    \int_0^{2\pi}
    \frac{\dd q_x}{2\pi}
    h_n
    \qty(L_y^{-1}\sum_{i_y=0}^{L_y-1}\cos \Delta_{q_x,i_y})
    \\
    \times 
    \min(\ell,2t|v_x(q_x)|).
    \label{eq:Sn_ITF_QP}
\end{multline}

In Fig.~\ref{fig:Sn_ITF}, we plot as a function of time the R\'enyi entanglement entropies in the quench from the inhomogeneous partially-filled product state with the inhomogeneous angle $\theta_{i_y}=2\pi i_y/L_y$. The plot shows that the expression of Eq.~\eqref{eq:Sn_ITF_SP_limit} (solid lines) agrees well with the exact values (symbols) of the entropies obtained by numerically with Eq.~\eqref{eq:S_n=tr_ln_Gamma}, whereas the prediction of the quasiparticle picture in Eq.\,\eqref{eq:Sn_ITF_QP} (dashed curves) does not match the correct result. 
\par 
This discrepancy means that, as in the case of the crossed dimer state, the entanglement entropies derived from the GGEs associated with $H$ and $H^X$ are different, and the correct one is the latter. 
In the present case, one finds that the GGE that reproduces the saturation value of the entanglement entropy in Eq.\,\eqref{eq:Sn_ITF_t_limit} is built with the charges 
\begin{align}
    \hat n_{q_x,i_y} = d_{q_x,i_y}^\dag d_{q_x,i_y},
\end{align}
which commute with $H^X$ but not with $H$. 
\par 
Repeating the derivation of Eq.\,\eqref{eq:Sn(rho_GGE,A)} for the current case, we find that the entropy of the GGE built with $\{\hat n_{q_x,i_y}\}$ is
\begin{align}
    S_n
    =
    \sum_{i_y=0}^{L_y-1}
    \int_0^{2\pi}\frac{\dd q_x}{2\pi}
    h_n(2n_{q_x,i_y}-1),
    \label{eq:S_n(rho_GGEXA)_ITF}
\end{align}
where $n_{q_x,i_y}=\bra{\psi_0} \hat n_{q_x,i_y}\ket{\psi_0}$. 
For the state $\ket{\mathrm{IPF}}$, this expectation value is  
\begin{align}
    n_{q_x,i_y}
    = 
    \frac{1-\cos \Delta_{q_x,i_y}}{2}. 
\end{align}
Plugging it into Eq.\,\eqref{eq:S_n(rho_GGEXA)_ITF}, we indeed obtain Eq.\,\eqref{eq:Sn_ITF_t_limit}.

\end{document}